\documentclass[letter, a4paper]{IEEEtran}
\pdfoutput=1
\usepackage{amsmath,amsfonts}
\usepackage{setspace}
\usepackage{algorithmic}
\usepackage{algorithm}
\usepackage{array}
\usepackage{textcomp}
\usepackage{stfloats}
\usepackage{url}
\usepackage{verbatim}
\usepackage{graphicx}
\usepackage{cite}
\usepackage{framed}
\usepackage{bm}
\usepackage{xcolor}
\usepackage{xpatch}
\usepackage{multirow}
\usepackage{diagbox}  
\usepackage{xcolor}  
\usepackage{pifont}
\usepackage{stfloats}  
\usepackage{graphicx} 
\usepackage{epsfig}
\usepackage{epstopdf} 

\ifCLASSOPTIONcompsoc
\usepackage[caption=false,font=normalsize,labelfont=sf,textfont=sf]{\part{title}subfig}
\else
\usepackage{subfigure}
\fi
\usepackage{graphicx,amsmath,amssymb,amsfonts}
\allowdisplaybreaks[3]
\usepackage{caption}
\usepackage{hyperref}     
\hypersetup{colorlinks=true}    
\usepackage{xcolor}
\usepackage[T1]{fontenc} 

\newtheorem{theorem}{\textbf{Theorem}}

\newtheorem{lemma}{\textbf{Lemma}}

\newtheorem{corollary}{\textbf{Corollary}}

\makeatletter

\newcommand{\Rmnum}[1]{\expandafter\@slowromancap\romannumeral #1@}

\makeatother

\begin{document}
	\captionsetup{font={small}}
	\bstctlcite{reference:BSTcontrol}
	
	\title{\fontsize{21 pt}{\baselineskip}\selectfont Energy-Efficient Robust Beamforming for Multi-Functional RIS-Aided Wireless Communication under Imperfect CSI}
	\author{
		Ailing~Zheng,
		Wanli~Ni, 
		Wen~Wang,
		Hui~Tian,
		and Chau~Yuen
		\thanks{The work of Hui Tian was supported by the Natural Science Foundation of Shandong Province under Grant No. ZR2021LZH010. The work of Chau Yuen was supported by the Ministry of Education, Singapore, under its MOE Tier 2 (Award number MOE-T2EP50220-0019), and National Research Foundation, Singapore and Infocomm Media Development Authority under its Future Communications Research \& Development Programme FCPNTU-RG-2024-025. The review of this article was coordinated by Prof. Mark Flanagan. (Corresponding author: Hui Tian.)}
		\thanks{A. Zheng and H. Tian are with the State Key Laboratory of Networking and Switching Technology, Beijing University of Posts and Telecommunications, Beijing 100876, China (e-mail: ailing.zheng@bupt.edu.cn, tianhui@bupt.edu.cn).} 
		\thanks{W. Ni is with the Department of Electronic Engineering, Tsinghua University, Beijing 100084, China (e-mail: niwanli@tsinghua.edu.cn).}
		\thanks{W. Wang is with the Pervasive Communications Center, Purple Mountain Laboratories, Nanjing 211111, China, and also with the School of Information Science and Engineering, and the National Mobile Communications Research Laboratory, Southeast University, Nanjing 210096, China (email: wangwen@pmlabs.com.cn).}
		\thanks{C. Yuen is with the school of Electrical and Electronic Engineering, Nanyang Technological University, Singapore (e-mail: chau.yuen@ntu.edu.sg).}
	}
	\maketitle
	\begin{abstract}
		The robust beamforming design in multi-functional reconfigurable intelligent surface (MF-RIS) assisted wireless networks is investigated in this work, where the MF-RIS supports signal reflection, refraction, and amplification to address the double-fading attenuation and half-space coverage issues faced by traditional RISs. Specifically, we aim to maximize the system energy efficiency by jointly optimizing the transmit beamforming vector and MF-RIS coefficients in the case of imperfect channel state information (CSI). We first leverage the S-procedure and Bernstein-Type Inequality approaches to transform the formulated problem into tractable forms in the bounded and statistical CSI error cases, respectively. Then, we optimize the MF-RIS coefficients and the transmit beamforming vector alternately by adopting an alternating optimization framework, under the quality of service constraint for the bounded CSI error model and the rate outage probability constraint for the statistical CSI error model. Simulation results demonstrate the significant performance improvement of MF-RIS compared to benchmark schemes.
		In addition, it is revealed that the cumulative CSI error caused by increasing the number of RIS elements is larger than that caused by increasing the number of transmit antennas.
	\end{abstract}
	
	\begin{IEEEkeywords}
		Multi-functional RIS, energy efficiency maximization, robust beamforming, imperfect CSI.
	\end{IEEEkeywords}
	
	\section{Introduction}
	The forthcoming sixth-generation (6G) wireless network demands high spectral efficiency and massive connectivity to meet the exponential growth of wireless traffic \cite{K2022-6G}. Various advanced technologies, including non-orthogonal multiple access (NOMA), multiple-input multiple-output (MIMO), and millimeter wave (mmWave) communications, have been exploited to support low latency, high data rates, and massive device access. Nevertheless, in most cases, achieving high spectrum efficiency often entails a significant increase in energy consumption \cite{Huang2019-EE}. To balance between throughput performance and power consumption, energy efficiency (EE) has been recognized as an important performance metric to measure whether 6G communications meet green and sustainability requirements \cite{Ihsan2022-EE}. Therefore, innovative technologies to support spectrum- and energy-efficient communications are urgent.
	
	Benefiting from the fabrication of programmable meta-surfaces, reconfigurable intelligent surface (RIS) provides a significant and viable solution to boost the spectral efficiency, EE, and signal coverage in 6G networks. Generally, an RIS is a planar array of a large number of low-cost elements that can be dynamically adjusted to reconfigure electromagnetic waves \cite{Wu2019-RIS}. Thus, an RIS is capable of altering the amplitude and phase of incident signals by intelligently modulating the phase shift of each element, subsequently redirecting the signals towards the desired direction \cite{Liu2021-RIS}. This capability enables RIS to tailor the wireless environment to fulfill specific network demands.
	Since the RIS only reflects the incident signals without involving the decoding and regeneration processes, it boasts significantly lower power consumption and hardware cost compared to relays \cite{Huang2019-RIS}. Therefore, the RIS has been recognized as a potential technology to tackle the problem of high energy consumption in wireless communications.
	
	However, since the single-functional RIS (SF-RIS) only reflects or refracts the impinging signals, the resulting geographical confinement limits the flexibility and performance improvement of the SF-RIS deployment.
	Specifically, the SF-RIS provides services only for users located on the reflection side, resulting in a half-space coverage. 
	To overcome this limitation, the dual-functional RIS (DF-RIS), e.g., simultaneously transmitting and reflecting (STAR)-RIS \cite{Xu2021-STAR} or intelligent omni-surface (IOS) \cite{Zhang2022-IOS} has been proposed to support signal reflection and refraction simultaneously, thus building a full-space smart radio environment. 
	Nevertheless, the signals relayed by DF-RISs undergo the cascaded channel of the transmitter-RIS-receiver. The resulting double-fading attenuation seriously weakens the signal strength and degrades the advantages of RIS deployment in wireless networks \cite{Najafi2021}. Although deploying a large RIS with numerous passive elements may compensate for the high path-loss, the corresponding high control overhead and power consumption are inevitable. 
	To alleviate the adverse effect faced by DF-RISs, active RISs with signal reflection and amplification are introduced to transcend the inherent physical limits of the double fading effect \cite{Long2021-activeRIS}. Specifically, each element integrated with negative resistance components, e.g., tuning diodes, can convert the direct current (DC) bias power to radio frequency (RF) power, thus amplifying the received signal without significantly compromising the low power budget requirements \cite{J2014}. Therefore, the active RIS can alleviate the double-fading attenuation while ensuring low power consumption. 
	
	Furthermore, to support full-space coverage and overcome the double-fading attenuation at the same time, we introduced the multi-functional RIS (MF-RIS) to reflect, refract, and amplify the incident signals simultaneously \cite{Ni2024}. 
	However, existing works on MF-RIS are conducted in the case of perfect channel state information (CSI) case \cite{Zheng2023,Wang2023-RIS,Zhang2023-ISAC}, which is challenging to obtain in practice. Moreover, due to the integration of amplifiers, both the received desired signal and thermal noise are amplified at MF-RIS \cite{Zhou2024CSI}, thus the cascaded CSI of the transmitter-RIS-receiver exploited in passive RISs can not be utilized for beamforming design at the MF-RIS \cite{Wang2020CSI}. In addition, since MF-RISs lack RF chains, they can not perform channel estimation on two individual links. Therefore, it is essential to design robust beamforming strategies for MF-RIS assisted networks.

	\vspace{-2mm}
	\subsection{Related Works}
	As the RIS has gained growing research attention, several related aspects including EE improvement via RIS, robust beamforming for RIS, and MF-RIS-aided wireless communications, have been extensively investigated in the literature.
	
	\subsubsection{EE improvement via RIS}
	Due to the nearly-passive feature, various research activities have been devoted in RIS-empowered networks for EE improvement \cite{Yang2022-RIS,Ma2023-RIS,Fang2020-RIS,Fang2022-STAR,Guo2023-STAR,Ma2023-ARIS,Fotock2023-ARIS,Fotock-2024-ARIS}. Specifically, the authors of \cite{Fang2020-RIS} maximized EE in an SF-RIS-aided NOMA network and proved that the SF-RIS can greatly improve the EE performance. Then, the authors of \cite{Ma2023-RIS} proposed an SF-RIS-enhanced simultaneous wireless information and power transfer (SWIPT) system built on an electromagnetic-compliant framework, where the EE was maximized under the impedance parameters of SF-RIS elements constraint. Additionally, in \cite{Yang2022-RIS}, the EE maximization problem in a wireless network with distributed SF-RISs was investigated by optimizing RIS on-off status, which revealed that the multi-RIS scheme obtained a higher EE performance than the single-RIS scheme.
	Furthermore, since the STAR-RIS supports full-space coverage to serve users, the authors of \cite{Fang2022-STAR} investigated the energy-efficient resource allocation scheme to maximize EE in a STAR-RIS-aided MIMO-NOMA network, which revealed that NOMA can achieve superior EE performance than orthogonal multiple access (OMA) scheme. Besides, in \cite{Guo2023-STAR}, the system EE was maximized by exploiting a deep deterministic policy gradient (DDPG)-based algorithm in STAR-RIS-aided NOMA networks. 
	In addition, to alleviate the double-fading attenuation issue, the active RIS has been extensively studied recently. For instance,
	the EE comparison between active RISs and passive RISs was provided in \cite{Fotock2023-ARIS}, which revealed that whether active RISs or passive RISs is more energy efficient depends on the energy consumption of each active RIS element. 
	To proceed, the authors of \cite{Ma2023-ARIS} maximized EE in an active RIS-aided multiuser system, and illustrated that the active RIS outperforms half/full duplex relays under the same power budgets. Additionally, in \cite{Fotock-2024-ARIS}, the global EE (GEE) was optimized in active RIS-aided systems with global reflection constraints, which revealed that active RISs provide a higher GEE than nearly-passive RISs only if the additional power consumption is not larger than the hardware power consumption of nearly-passive RISs. These studies demonstrated that the RIS has great potential in improving system EE performance.
	
	\subsubsection{Robust beamforming for RIS}
	Considering that the RIS is nearly-passive without bulky RF chains, the perfect CSI is difficult to acquire. Thus, a surge of research has focused on robust beamforming designs for RIS-aided networks \cite{Zhou2024CSI,Zhou2020CSI,Peng2022-RIS,Mahdi2023-RIS,Wang2022CSI,Li2021-STAR,Lyu2023-ARIS,Mahdi2024-RIS,Li2023-ARIS}.  
	Particularly, the authors of \cite{Zhou2020CSI} investigated the transmit power minimization problem under imperfect CSI conditions, which showed that whether to deploy the RIS depends on the level of the CSI error.
	In addition to imperfect CSI, in \cite{Peng2022-RIS}, a robust beamforming algorithm with hardware defects was investigated, which demonstrated the effectiveness of the proposed algorithm. Furthermore, the authors of \cite{Mahdi2023-RIS} studied a robust two-timescale transmission design in SF-RIS assisted cell-free massive MIMO networks. The effect of pilot contamination on channel estimation was studied, and the expression for spectral efficiency was then obtained in closed form. Since the STAR-RIS supports both signal reflection and refraction, the CSI acquisition is more challenging.
	In \cite{Wang2022CSI}, the secure transmission in a STAR-RIS-aided NOMA system with imperfect CSI was investigated, which revealed that the size of RIS deployed in real-world scenarios depends on the CSI uncertainty.
	Furthermore, the authors of \cite{Li2021-STAR} theoretically analyzed the ergodic sum-rate of the STAR-RIS-assisted NOMA scheme under estimation errors and hardware impairments, which revealed that both the channel estimation and the ergodic sum-rate have performance floor at high transmit power regions caused by transceiver hardware impairments.
	Owing to the signal amplification achieved by amplifiers, the robust beamforming design for active RIS is necessary. In particular, the authors of \cite{Zhou2024CSI} derived closed-form expressions of the average data rate and the RIS transmit power for transceiver design based on partial CSI in active RIS-aided systems, in which a rate outage constraint beamforming problem was formulated and solved using the Bernstein-Type inequality. 
	Moreover, the authors of \cite{Lyu2023-ARIS} proposed a robust secure communication scheme in active RIS-assisted symbiotic radio networks, which demonstrated that the active RIS can save more energy compared to passive RISs.
	In \cite{Mahdi2024-RIS}, the authors obtained the closed-form expressions for uplink spectral efficiency and amplitude gain in active RIS-assisted MIMO networks in the case of imperfect CSI.
	In addition, the outage probability and ergodic channel capacity were derived in closed-form with channel estimation errors in \cite{Li2023-ARIS}, which proved that incorporating amplifiers in RIS is comparable to boosting the transmit power, thereby enhancing the system performance.   
	
	\subsubsection{MF-RIS-aided wireless communications}
	By reflecting, refracting, and amplifying the impinging signal simultaneously, the MF-RIS can address both the double-fading attenuation and half-space coverage issues. Therefore, research on MF-RIS has been carried out recently \cite{Wang2023-RIS,Guo2023-RIS,Zhang2023-ISAC,Yan2023-RIS,Zheng2023-MF-RIS,Xie2023-MF-RIS}.
	Specifically, the operation principles of the physical architecture and equivalent circuit model of each MF-RIS element were outlined in \cite{Wang2023-RIS}, illuminating its operation mechanism.
	The authors of \cite{Guo2023-RIS} utilized the MF-RIS to promote secrecy performance by maximizing sum secrecy rate and demonstrated the superiority of MF-RIS.
	In \cite{Zhang2023-ISAC}, the sum rate was maximized in an MF-RIS-aided integrated sensing and communications (ISAC) system under the constraints of sensing performance, and the radar receiving beamforming vector was derived in closed-form. Moreover, by switching each MF-RIS element between reflection and refraction modes, the authors of \cite{Yan2023-RIS} formulated a sum rate maximization problem in MF-RIS-assisted networks.
	In addition, in \cite{Zheng2023-MF-RIS}, three operating protocols in MF-RIS-aided NOMA networks with discrete phase shifts was proposed.
	In \cite{Xie2023-MF-RIS}, the authors derived asymptotic expressions for the outage probability, considering both perfect and imperfect successive interference cancellation (SIC) cases for users, and evaluated the overall throughput of MF-RIS-assisted networks. 
	
	\begin{table*}[t]
		\centering
		\caption{Comparison of related papers with this work}
		\scalebox{0.85}{
			\begin{tabular}{|c|c|c|c|c|c|c|c|c|} 
				\hline 
				\diagbox{Properties}{References}   & \cite{Fang2020-RIS,Yang2022-RIS,Ma2023-RIS} & \cite{Zhou2020CSI,Peng2022-RIS,Mahdi2023-RIS} & \cite{Fang2022-STAR,Guo2023-STAR}  &  \cite{Wang2022CSI} & \cite{Ma2023-ARIS,Fotock2023-ARIS,Fotock-2024-ARIS}  & \cite{Zhou2024CSI,Lyu2023-ARIS,Li2023-ARIS,Mahdi2024-RIS} & \cite{Wang2023-RIS,Guo2023-RIS,Zhang2023-ISAC,Yan2023-RIS,Zheng2023-MF-RIS} & \bf{This work} \\ 
				\hline 
				Full-space coverage   &  &  & \textcolor{black}{\ding{51}} & \textcolor{black}{\ding{51}} &  &  & \textcolor{black}{\ding{51}} & \textcolor{black}{\ding{51}} \\
				\hline
				Double-fading mitigation  &  &  &  &  & \textcolor{black}{\ding{51}} & \textcolor{black}{\ding{51}} & \textcolor{black}{\ding{51}} & \textcolor{black}{\ding{51}} \\
				\hline
				EE performance   & \textcolor{black}{\ding{51}} &  & \textcolor{black}{\ding{51}} & \textcolor{black}{\ding{51}} & \textcolor{black}{\ding{51}} &  &  & \textcolor{black}{\ding{51}} \\
				\hline
				Imperfect CSI   &  & \textcolor{black}{\ding{51}} &  & \textcolor{black}{\ding{51}}  && \textcolor{black}{\ding{51}}  &  & \textcolor{black}{\ding{51}} \\
				\hline
				RIS architecture   & \multicolumn{2}{c}{{SF-RIS}} & \multicolumn{2}{|c|}{{STAR-RIS}}  & \multicolumn{2}{c}{{Active RIS}} & \multicolumn{2}{|c|}{{MF-RIS}}  \\
				\hline
		\end{tabular}}
		\label{Com}
	\end{table*}
	Table \ref{Com} compares this work with other representative studies. It is observed that compared to other RIS types, the MF-RIS can achieve full-space coverage while simultaneously mitigating double fading. However, although several papers have investigated the MF-RIS, they focused on the system throughput maximization or the transmit power consumption minimization.
	Since the MF-RIS requires additional energy for signal amplification, the EE performance to boost the tradeoff between the throughput performance and power consumption is necessary to investigate, which is missing in the current research. Additionally, due to the inevitably amplified thermal noise at the MF-RIS, the perfect CSI is hard to obtain in actual scenarios. The works about MF-RIS assuming perfect CSI may inevitably result in undesired performance degradation \cite{Wang2023-RIS,Guo2023-RIS,Zhang2023-ISAC,Yan2023-RIS,Zheng2023-MF-RIS}. Therefore, it motivates us to study the EE performance of MF-RIS with imperfect CSI to provide a more general performance.
	
	\subsection{Contributions and Organization}
	Based on the above discussion, in this work, we study the robust beamforming scheme in an MF-RIS-aided network under imperfect CSI conditions. Specifically, we investigate an EE maximization problem for the bounded and the statistical CSI error models. The main contributions are summarized as follows:
	
	\begin{itemize}
		\item
		We propose a robust beamforming design for an MF-RIS-aided multi-user system to enhance signal strength and support full-space coverage in this paper. Specifically, we formulate an EE maximization problem to balance between throughout performance and power consumption, where the transmit beamforming vector and the MF-RIS coefficients are optimized jointly under different CSI error models.
		\item
		For the bounded CSI error model, we maximize the system EE while ensuring quality of service (QoS) requirements for users and power consumption constraints of the BS and the MF-RIS. Particularly, we first utilize the S-procedure and general sign-definiteness techniques to approximate semi-infinite inequality constraints and then transform the obtained problem to avoid an infinite number of constraints. Then, we adopt an alternating optimization (AO) framework to optimize the transmit beamforming vector and the MF-RIS coefficients alternately, where the sequential convex approximation (SCA) and penalty convex-concave procedure (PCCP) methods are exploited in each subproblem.
		
		\item
		For the statistical CSI error model, we formulate an EE maximization problem subject to the rate outage probability constraints. Initially, we derive an analytical expression for the average MF-RIS amplification power constraint. Then, we apply the Bernstein-Type Inequality to transform the rate outage probability constraint into a tractable form. Finally, the semi-definite relaxation (SDR) is utilized to iteratively tackle the transmit beamforming vector and the MF-RIS coefficients sub-problems, in which the sequential rank-one constraint relaxation (SROCR) method is adopted to relax the rank-one constraint.
		\item
		Numerical results demonstrate the convergence and effectiveness of the proposed algorithms. Besides, the MF-RIS in the case of perfect CSI has only 4\% EE performance improvement compared to imperfect CSI, which reveals the robustness of the proposed algorithm.
		Moreover, the EE performance of the statistical CSI error model outperforms that of the bounded CSI error model, exhibiting a 8\% performance gap in the MF-RIS scheme. Furthermore, the MF-RIS outperforms other schemes, among which it enjoys 24\% , 51\%, and 86\% higher EE than STAR-RIS, active RIS, and SF-RIS schemes, respectively. In addition, due to the accumulated CSI errors, the feasibility rate of the MF-RIS scheme decreases with the increase of  the number of RIS elements and transmit antennas.
	\end{itemize}
	
	The rest of this paper is organized as follows. In Section \ref{System Model}, we present the system model and CSI error models for the MF-RIS-aided communication system. Sections \ref{bounded} and \ref{statistical} delve into the robust beamforming designs for the bounded and the statistical CSI error models, respectively. The convergence behavior and the computational complexity of the proposed algorithm are analyzed in Section \ref{Convergence}. Finally, numerical results and conclusions are provided in Sections \ref{Simulation} and \ref{Conclusion}, respectively.
	
	\emph{Notations:} Vectors and matrices are denoted by boldface lowercase letters and boldface uppercase letters, respectively. $\mathbb{H}^{N}$ represents the set of all $N$-dimensional complex Hermitian matrices. $\mathbf{X}^{\ast}$, $\mathbf{X}^{\mathrm{T}}$, $\mathbf{X}^{\mathrm{H}}$, and $\Vert \mathbf{X} \Vert_F$ denote the conjugate, transpose, Hermitian, Frobenius norm of matrix $\mathbf{X}$, respectively. $\mathbf{X} \succeq \mathbf{0}$ indicates that matrix $\mathbf{X}$ is positive semi-definite. $\Vert \mathbf{x} \Vert_2$ denotes 2-norm of vector $\mathbf{x}$. $\mathrm{diag}(\mathbf{x})$ is a diagonal matrix with the entries of $\mathbf{x}$ on its main diagonal. $[\mathbf{x}]_m$ denotes the $m$-th element of vector $\mathbf{x}$ and the submatrix $[\mathbf{X}]_{1:m,1:m}$ denotes the first $m$ columns and $m$ rows elements of matrix $\mathbf{X}$.
	$\Re\{\cdot\}$, and $\mathrm{Tr}\{\cdot\}$,  denote the real part, trace, respectively. $\otimes$ and $\cdot$ represent the Kronecker product and the dot product, respectively. Additionally, $\mathbb{C}$ and $\mathbb{R}$ denote the complex field and real field, respectively.
	
	\section{System Model} \label{System Model}
	\subsection{MF-RIS-Aided Communications}
\begin{figure}[t]
		\centering
		\includegraphics[width=3.4 in]{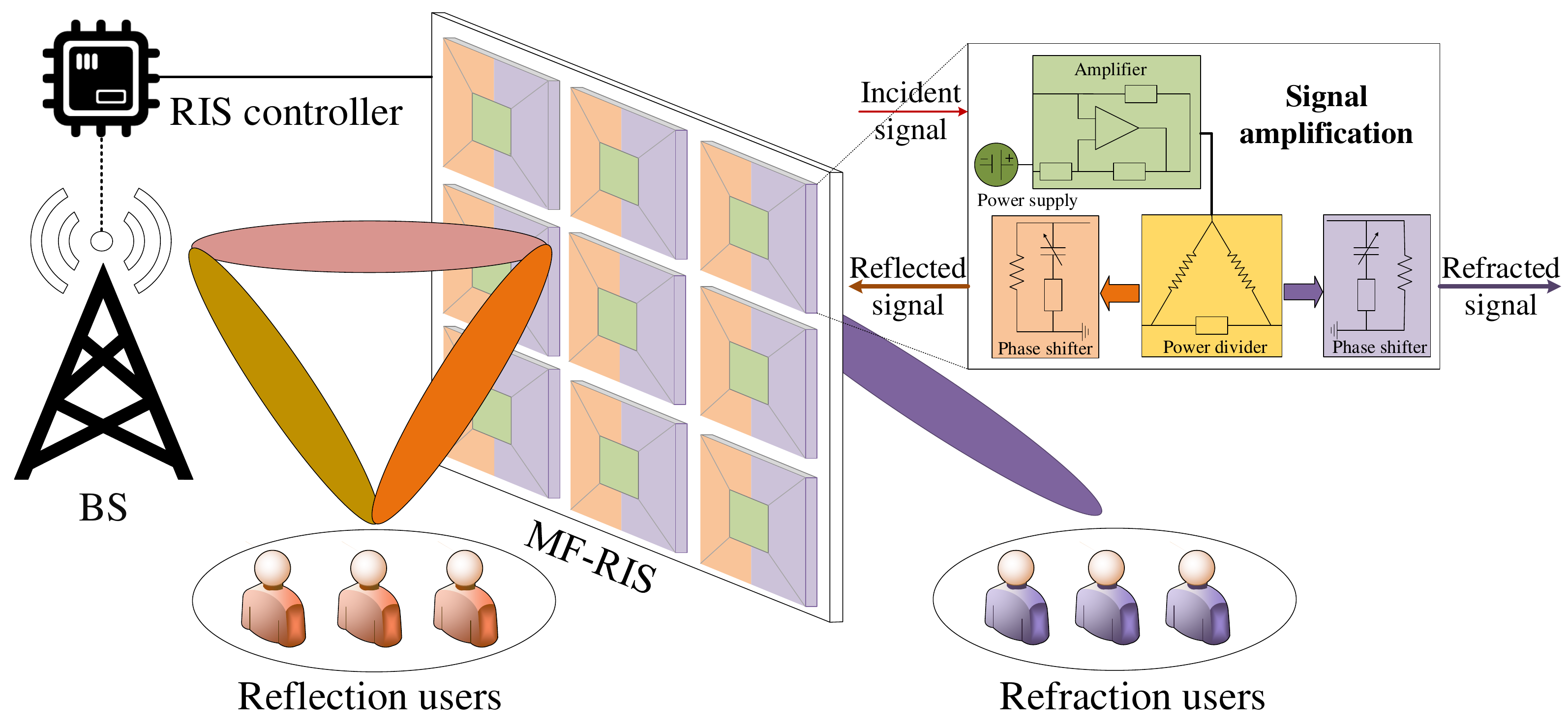}
		\caption{An MF-RIS-aided multiuser communication system.}
		\label{system_model}
		\vspace{-5mm}
	\end{figure}

	As shown in Fig. \ref{system_model}, we investigate an MF-RIS-aided multiuser multiple-input single-output (MISO) wireless network, where an MF-RIS consisting of $M$ elements is utilized to assist the transmission from an $N$-antenna BS to $K$ single-antenna users.
	The user $k$ in space $c$ is denoted by $U_{c,k}$. We have $\mathcal{K}_c = \{1,2,\cdots, K_c\}$  with the number of all users $K = K_r + K_t$, where $c \in \mathcal{C} = \{t,r\}$ represents reflection ($c = r$) and refraction ($c =  t$) space, respectively. 
	Denote $\mathbf{u}_c = [\sqrt{\beta_{1}^c}e^{j\theta_1^c}, \sqrt{\beta_2^c}e^{j\theta_2^c}, \cdots, \sqrt{\beta_M^c}e^{j\theta_M^c}]^{\mathrm{T}} \in  \mathbb{C}^{M \times 1}$, then we have $\boldsymbol{\Theta}_c=\mathrm{diag}(\mathbf{u}_c)$ as the refraction or reflection matrix of MF-RIS, where $\beta_m^c \in [0,\beta_{\rm max}]$ and $\theta_m^c \in [0,2\pi)$ represent the amplitude and phase shift of the $m$-th element, respectively. Here the amplitude coefficient $\beta_m^c$ satisfies $\beta_m^r + \beta_m^t \le \beta_{\rm max}$ with $\beta_{\rm max} \ge 1$.
	The channels from the BS to the MF-RIS, the BS to user $U_{c,k}$, and the MF-RIS to user $U_{c,k}$ are given by $\mathbf{G} \! \in \! \mathbb{C}^{M \times N}$, $\mathbf{h}_{c,k} \in \mathbb{C}^{1 \times N}$, and $\mathbf{f}_{c,k} \in \mathbb{C}^{1 \times M}$, respectively {\footnote{The partial CSI of all individual links can be obtained at the BS by using existing channel estimation methods such as that in \cite{Shtaiwi-2021}.}}. To simplify the model, we assume that the self-interference (SI) at the MF-RIS can be perfectly suppressed \cite{Zhang2023}.
	Then, the signal received at user $U_{c,k}$ is given by
	\setlength{\abovedisplayskip}{1pt}
	\setlength{\belowdisplayskip}{2pt}
	\begin{eqnarray}
		&& \!\!\!\!\!\!\!\!\!\!\!\!\!\!\!\!\!\!\!\! y_{c,k}\!=\!(\mathbf{h}_{c,k}\!+\!\mathbf{f}_{c,k}\mathbf{\Theta}_c\mathbf{G}) \sum\limits_{c\in \mathcal{C}}\sum\limits_{k=1}^{K_c}\mathbf{w}_{c,k}s_{c,k}\!+\!\mathbf{f}_{c,k}\mathbf{\Theta}_c\mathbf{z}_1\!+\! z_2,
	\end{eqnarray}
	where $\mathbf{w}_{c,k} \! \in \! \mathbb{C}^{N \times 1}$ denotes the transmit beamforming vector, and $s_{c,k} \sim \mathcal{CN}(0,1)$ is the information-bearing symbol for user $U_{c,k}$. Here, $\mathbf{z}_1 \sim \mathcal{CN}(\mathbf{0},\sigma_1^2\mathbf{I}_M)$ and $z_2 \sim \mathcal{CN}(0,\sigma_2^2)$ are the thermal noise at the MF-RIS and user $U_{c,k}$, respectively.
	
	Let $\bar{\mathbf{h}}_{c,k}\!=\! \mathbf{h}_{c,k}+\mathbf{f}_{c,k}\mathbf{\Theta}_c\mathbf{G}\! \in\! \mathbb{C}^{1 \times N}$, and  $\mathbf{W}_{c,-k}=[\mathbf{w}_{c,1}, \mathbf{w}_{c,2}, \ldots, \mathbf{w}_{c,k-1}, \mathbf{w}_{c,k+1}, \ldots, \mathbf{w}_{c,K_c}, \mathbf{w}_{\bar{c},1}, \ldots, \mathbf{w}_{\bar{c},K_{\bar{c}}}]$, here $\bar{c} = t$ when $c = r$, , and vice versa.
	The signal-to-interference-plus-noise-ratio (SINR) of user $U_{c,k}$ is given by
	\setlength{\abovedisplayskip}{-1pt}
	\setlength{\belowdisplayskip}{1pt}
	\begin{eqnarray}
		&&\!\!\!\!\!\!\!\!\!\!\!\!\!\!\!\!\!\!\! \gamma_{c,k} \!= \!\frac{\vert \bar{\mathbf{h}}_{c,k}\mathbf{w}_{c,k} \vert^2}
		{\Vert \bar{\mathbf{h}}_{c,k}\mathbf{W}_{c,-k} \Vert_2^2 + \Vert \mathbf{f}_{c,k}\mathbf{\Theta}_c \Vert_2^2 \sigma_1^2 + \sigma_2^2},~\forall c,~\forall k.
	\end{eqnarray}
	Then, the corresponding communication rate of user $U_{c,k}$ is $R_{c,k}=\log_2(1+\mathrm{\gamma}_{c,k})$.
	
	For the total power consumption \footnote{	
		In this work, we assume that the channel condition is sufficiently stable that the pilot overhead has relatively low influence on the EE. Furthermore, since the focus of this work lies in the robust beamforming design of MF-RIS-aided systems, detailed channel estimation methods and their specific effects on EE are ignored for simplicity.
		}, we have 
	\setlength{\abovedisplayskip}{1pt}
	\setlength{\belowdisplayskip}{1pt}
	\begin{align}
		P_{\mathrm{total}} = & \xi \sum\nolimits_{c}\sum\nolimits_{k=1}^{K_c} \Vert \mathbf{w}_{c,k} \Vert^2 + \zeta P_{\mathrm{RIS}} \nonumber \\
		& +  KP_U +  P_{\mathrm{BS}} + 2M(P_{\mathrm{PS}} + P_{\mathrm{PA}}), 
	\end{align}
	where $\xi$ and $\zeta$ are the inverse of energy conversion coefficients at the BS and the MF-RIS, respectively. $P_{\mathrm{RIS}}$ represents the energy consumption at the MF-RIS with $P_{\mathrm{RIS}} = \sum\nolimits_{c}(\Vert \boldsymbol{\Theta}_c \mathbf{G}\mathbf{W}  \Vert_F^2 + \Vert \boldsymbol{\Theta}_c \Vert_F ^2 \sigma_1^2)$, where $\mathbf{W}=[\mathbf{w}_{c,1}, \mathbf{w}_{c,2}, \ldots, \mathbf{w}_{c,K_c}, \mathbf{w}_{\bar{c},1}, \ldots, \mathbf{w}_{\bar{c},K_{\bar{c}}}]$. $P_U$ and $P_{\mathrm{BS}}$ denote the static power consumed at each user and the BS with $P_{\mathrm{BS}}=P_{\rm S}+NP_{\rm RF}$, where $P_{\rm S}$ is the constant power required for BS operation and $P_{\rm RF}$ denotes the power necessary to operate the circuit components attached to each antenna.
	Moreover, $P_{\mathrm{PS}}$ and $P_{\mathrm{PA}}$ are the power consumed by each phase shifter and the DC biasing power consumed by each amplifier, respectively. 
	
	\vspace{-2mm}
	\subsection{Channel Uncertainty Model}
	Although the case of perfect CSI provides an upper bound on performance, the CSI error in practical scenarios is inevitable due to channel estimation errors, channel feedback delays, and quantization errors. To study the effect, we exploit two commonly-used CSI error models, i.e., the bounded CSI error model and the statistical CSI error model, to characterize the CSI uncertainty of the corresponding channels. Both CSI error models give insight into how the system performs under uncertainty, which is crucial for designing robust communication systems that perform well even with imperfect information.
	The channel uncertainty is  modeled as follows:
	\begin{subequations}
		\begin{align}	
			&\mathbf{h}_{c,k}=\mathbf{\widehat{h}}_{c,k}+\Delta\mathbf{h}_{c,k}, ~\mathbf{f}_{c,k}=\mathbf{\widehat{f}}_{c,k}+\Delta\mathbf{f}_{c,k},~\forall c,~\forall k,\\
			&\mathbf{F}_{c,k}=\mathbf{\widehat{F}}_{c,k}+\Delta\mathbf{F}_{c,k},  
			~\mathbf{G}=\mathbf{\widehat{G}}+\Delta\mathbf{G},~\forall c,~ \forall k,
		\end{align}
	\end{subequations}
	where $\mathbf{F}_{c,k}=\mathrm{diag}(\mathbf{f}_{c,k})\mathbf{G}$ is the cascaded channel from the BS to user $U_{c,k}$. $\mathbf{\widehat{h}}_{c,k}$, $\mathbf{\widehat{f}}_{c,k}$, $\mathbf{\widehat{F}}_{c,k}$, and $\mathbf{\widehat{G}}$ are the estimated CSI of $\mathbf{h}_{c,k}$, $\mathbf{f}_{c,k}$, $\mathbf{F}_{c,k}$, and $\mathbf{G}$ known at the BS, respectively. $\Delta\mathbf{h}_{c,k}$, $\Delta\mathbf{f}_{c,k}$, $\Delta\mathbf{F}_{c,k}$, and $\Delta\mathbf{G}$ denote the corresponding CSI estimation errors. In this work, we study two types of the CSI error model.
	
	\emph{1) Case 1 (Bounded CSI error \cite{Botros2007CSI}):} 
	In this model, we define an uncertainty radius to quantify the degree of CSI uncertainty, which bounds the maximum norm of all CSI estimation errors. The bounded CSI error model characterizes the channel quantization error, which is inherently confined to a bounded region and can be applied to address quantization feedback errors.
	The specific channels are given by:
	\begin{subequations}
		\label{bounded CSI}
		\begin{align}
			&\!\!\! \Lambda_{h,c,k}=\!\{\Delta\mathbf{h}_{c,k} \in \mathbb{C}^{1 \times N}: \Vert \Delta\mathbf{h}_{c,k} \Vert_2 \le \xi_{h,c,k}\}, \forall c,~\forall k, \\ 
			&\!\!\! \Lambda_{f,c,k}=\!\{\Delta\mathbf{f}_{c,k} \in \mathbb{C}^{1 \times M}: \Vert \Delta\mathbf{f}_{c,k} \Vert_2 \le \xi_{f,c,k}\}, \forall c, ~\forall k, \\
			\label{F_k1}
			&\!\!\! \Lambda_{F,c,k}\!=\!\{\Delta\mathbf{F}_{c,k} \! \in\!  \mathbb{C}^{M \times N}:\! \Vert \Delta\mathbf{F}_{c,k} \Vert_F \! \le \!  \xi_{F,c,k}\}, \forall c, ~\forall k, \\
			&\!\!\! \Lambda_{G}=\!\{\Delta\mathbf{G} \in \mathbb{C}^{M \times N}: \Vert \Delta\mathbf{G} \Vert_F \le \xi_{G}\}, 
		\end{align}
	\end{subequations}
	where $\xi_{h,c,k}$, $\xi_{f,c,k}$, $\xi_{F,c,k}$, and $\xi_{G}$ are the radii of the uncertainty regions known at the BS.
	
	\emph{2) Case 2 (Statistical CSI error \cite{Zhang2011CSI}):} In this model, we assume that each CSI estimation error adheres to the circularly symmetric complex Gaussian (CSCG) distribution, which is commonly used to characterize the CSI imperfection caused by the channel estimation error. 
	The specific channels are presented below:
	\begin{subequations}
		\label{statistical CSI}
		\begin{align}	
			& \Delta\mathbf{h}_{c,k} \sim \mathcal{CN}(\mathbf{0},\mathbf{\Sigma}_{h,c,k}), ~\mathbf{\Sigma}_{h,c,k} \succeq \mathbf{0}, ~\forall c,~\forall k, \\
			& \Delta\mathbf{f}_{c,k} \sim \mathcal{CN}(\mathbf{0},\mathbf{\Sigma}_{f,c,k}), ~\mathbf{\Sigma}_{f,c,k} \succeq \mathbf{0}, ~\forall c,~\forall k, \\
			& \mathrm{vec}({\Delta\mathbf{F}_{c,k}}) \sim \mathcal{CN}(\mathbf{0},\mathbf{\Sigma}_{F,c,k}), ~\mathbf{\Sigma}_{F,c,k} \succeq \mathbf{0}, ~\forall c,~\forall k, \\
			& \mathrm{vec}(\Delta\mathbf{G}) \sim \mathcal{CN}(\mathbf{0},\mathbf{\Sigma}_{G}), ~\mathbf{\Sigma}_{G} \succeq \mathbf{0}, 
		\end{align}
	\end{subequations}
	where $\mathbf{\Sigma}_{h,c,k} \in \mathbb{C}^{N \times N}, ~\mathbf{\Sigma}_{f,c,k} \in \mathbb{C}^{M \times M}, ~\mathbf{\Sigma}_{F,c,k} \in \mathbb{C}^{MN \times MN}$, and $\mathbf{\Sigma}_{G} \in \mathbb{C}^{MN \times MN}$ are positive semidefinite error covariance matrices.
	
	\section{Robust Beamforming for Bounded CSI Error} \label{bounded}
	In this section, we delve into the investigation of the EE maximization problem under the bounded CSI error model.
	\subsection{Problem Formulation}
	In this part, we study the robust beamforming design under the bounded CSI error model to maximize EE of the MF-RIS-aided multiuser system, where the MF-RIS coefficients and the transmit beamforming vector are jointly optimized. The formulated optimization problem for this model is given by
	\begin{subequations}
		\label{P0}
		\begin{eqnarray}
			& \!\!\!\!\!\!\!\!\!\!\!\!\!\!\!\!\! \max \limits_{\mathbf{w}_{c,k},\boldsymbol{\Theta}_c} 
			\label{P0-function}
			& \!\!\!\! \frac{\sum\nolimits_{c}\sum\nolimits_{k=1}^{K_c} R_{c,k}}{P_{\mathrm{total}}}  \\
			\label{P0-C-transmit power}  
			&\!\!\!\!\!\!\!\!\!\!\!\!\!\!\!\!\! \mathrm{s.t.} & \!\!\!\!  \sum\nolimits_{c}\sum\nolimits_{k=1}^{K_c} \Vert \mathbf{w}_{c,k} \Vert ^2 \leq P_{\mathrm{BS}}^{\mathrm{max}},\\
			\label{P0-C-amplification power} 
			&& \!\!\!\!\sum\nolimits_{c}(\Vert \boldsymbol{\Theta}_c \mathbf{G}\mathbf{W}  \Vert_F^2 + \Vert \boldsymbol{\Theta}_c \Vert_F ^2 \sigma_1^2) \! \leq \! P_{\mathrm{RIS}}^{\mathrm{max}},~(\ref{bounded CSI}),\\
			\label{P0-C-RIS-coefficients}
			&& \!\!\!\! \sum\nolimits_{c} \beta_m^c \le \beta_{\max},~\forall c,~\forall m, \\
			\label{P0-C-RIS}
			&&\!\!\!\! \beta_m^c \in [0,\beta_{\max}],~ \theta_m^c \in [0,2\pi),~\forall c, ~\forall m,\\
			\label{P0-C-R_min}
			&& \!\!\!\!  R_{c,k} \ge R_{c,k}^{\mathrm{min}}, ~(\ref{bounded CSI}), ~\forall c,~ \forall k,
		\end{eqnarray}
	\end{subequations}
	where $P_{\mathrm{BS}}^{\mathrm{max}}$ and $P_{\mathrm{RIS}}^{\mathrm{max}}$ are the transmit power and amplification power at the BS and the MF-RIS, respectively. Constraints (\ref{P0-C-amplification power}) and (\ref{P0-C-transmit power}) represent the power budget constraints at the MF-RIS and the BS, respectively. In addition, constraint (\ref{P0-C-R_min}) ensures that the minimum rate requirement $R_{c,k}^{\mathrm{min}}$ is satisfied at user $U_{c,k}$.
	
	Note that problem (\ref{P0}) is challenging to tackle directly due to the highly coupled variables $\mathbf{w}_{c,k}$ and $\boldsymbol{\Theta}_c$, and the non-convex objective function. Moreover, the CSI estimation error introduced in the related channels results in infinitely many non-convex constraints (\ref{P0-C-amplification power}) and (\ref{P0-C-R_min}), which is challenging to solve. Therefore, we introduce two useful lemmas to deal with it and apply an AO framework to solve problem (\ref{P0}) in the following.

	\vspace{-2mm}
	\subsection{Problem Transformation}
	We initially transform the original problem (\ref{P0}) into a tractable form.
	By introducing slack variables $\psi$ and $\varrho$, we rewrite problem (\ref{P0}) as \cite{Wang2022CSI}
	\begin{subequations}
		\label{P1}
		\begin{eqnarray}
			& \max \limits_{\mathbf{w}_{c,k},\boldsymbol{\Theta}_c, \psi, \varrho} 
			\label{P1-function}
			&  \psi  \\
			\label{P1-1} 
			&  \mathrm{s.t.} & \sum\nolimits_{c}\sum\nolimits_{k=1}^{K_c} R_{c,k} \ge \psi \varrho,\\
			\label{P1-P_t}
			&& P_{\mathrm{total}} \le \varrho,  ~(\ref{bounded CSI}), ~ (\rm{\ref{P0-C-transmit power}})-(\rm{\ref{P0-C-R_min}}).
		\end{eqnarray}
	\end{subequations}
	
	For constraint (\ref{P1-1}), we apply the convex upper bound approximation to tackle the non-convex term $\psi \varrho$. Thus, when $t=\varrho / \psi $, we have
	\begin{align}
		g(\psi,\varrho) = \frac{t}{2}(\psi)^2 + \frac{(\varrho)^2}{2t} \ge \psi\varrho,
	\end{align}	
	where $t \geq 0$ is updated by $t^{(\tau)} = \varrho^{(\tau-1)} / \psi^{(\tau-1)} $ in the $\tau$-th iteration.
	Then, by introducing a slack variable set $\mathbf{r} = \{r_{c,k}, \forall (c,~k)\}$ with $r_{c,k}=R_{c,k}$, problem (\ref{P1}) is reformulated as 
	\begin{subequations}
		\label{P2}
		\begin{eqnarray}
			&\!\!\!\!\!\!\!\!\!\!\!\!\!\!\!\!\! \max \limits_{\mathbf{w}_{c,k}, \boldsymbol{\Theta}_c, \mathbf{r}, \psi, \varrho} 
			\label{P2-function}
			&\!\!\!\!\!\!\!\!  \psi  \\
			\label{P2-1} 
			&\!\!\!\!\!\!\!\!\!\!\!\!\!\!\!\!\!  \mathrm{s.t.} &\!\!\!\!\!\!\!\!\!\! \sum\nolimits_{c}\sum\nolimits_{k=1}^{K} r_{c,k} \ge g(\psi,\varrho),\\
			\label{P2-r}
			&&\!\!\!\!\!\!\!\!\!\!  r_{c,k} \ge R_{c,k}^{\mathrm{min}},~\forall c, ~\forall k, \\ 
			\label{P2-P-t}
			&&\!\!\!\!\!\!\!\!\!\! P_{\mathrm{total}} \!\le\! \varrho,\! ~(\ref{bounded CSI}),\!~ R_{c,k} \!\ge\! r_{c,k}, \!~(\rm{\ref{P0-C-transmit power}})-(\rm{\ref{P0-C-RIS}}).
		\end{eqnarray}
	\end{subequations}
	
	To solve the semi-definite constraint $R_{c,k} \ge r_{c,k}$ in constraint (\ref{P2-P-t}), we introduce two auxiliary variable sets $\bm{\eta} = \{\eta_{c,k}, \forall (c,~k)\}$ and  $\bm{\alpha} = \{\alpha_{c,k}, \forall (c,~k)\}$, then we have
	
	\begin{subequations}
		\begin{align}
			\label{SINR1}
			& \vert \bar{\mathbf{h}}_{c,k}\mathbf{w}_{c,k} \vert^2 \ge \alpha_{c,k}, ~\Lambda_{h,c,k}, ~\Lambda_{F,c,k}, \\
			\label{SINR2}
			& \Vert \bar{\mathbf{h}}_{c,k}\mathbf{W}_{c,-k} \Vert_2^2 + \mathcal{A}_{c,k} + \sigma_2^2 \le \eta_{c,k}, ~\Lambda_{h,c,k}, ~\Lambda_{F,c,k}, \\
			\label{SINR3}
			& \mathcal{A}_{c,k} \ge \Vert \mathbf{f}_{c,k}\mathbf{\Theta}_c \Vert_2^2 \sigma_1^2, ~\Lambda_{f,c,k}, \\
			\label{SINR0}
			& \frac{\alpha_{c,k}}{\eta_{c,k}} \ge 2^{r_{c,k}} - 1.
		\end{align}
	\end{subequations}
	Based on the Taylor-series expansion, the feasible set obtained always lies within the feasible set of problem (\ref{P1}). Thus, (\ref{SINR0}) can be linearized as \cite{Xu2022}
	\begin{align}
		\label{SINR0-1}
		\left\{\begin{aligned}
			&\alpha_{c,k}\geq e^{x_{c,k}^{(1)}},x_{c,k}^{(1)}-x_{c,k}^{(2)}\geq x_{c,k}^{(3)},\\
			&\eta_{c,k}\leq e^{\bar{x}_{c,k}^{(2)}}(x_{c,k}^{(2)}-\bar{x}_{c,k}^{(2)}+1),\\
			&2^{r_{c,k}}-1 \leq e^{\bar{x}_{c,k}^{(3)}}(x_{c,k}^{(3)}-\bar{x}_{c,k}^{(3)}+1),
		\end{aligned}\right.
	\end{align}
	where $x_{c,k}^{(p)}~(p \in (1),(2),(3))$ denotes the slack variable. Here  $\bar{x}_{c,k}^{(2)}$ and $\bar{x}_{c,k}^{(3)}$ are the feasible solutions of ${x}_{c,k}^{(2)}$ and ${x}_{c,k}^{(3)}$ obtained in the previous iteration.
	
	For constraint (\ref{SINR1}), we have $\vert \bar{\mathbf{h}}_{c,k}\mathbf{w}_{c,k} \vert^2 = \vert (\mathbf{h}_{c,k}+\mathbf{u}_{c,k}^{\mathrm{T}}\mathbf{F}_{c,k})\mathbf{w}_{c,k} \vert^2$. 
	Owing to the continuity of CSI uncertainty sets, the above constraints (\ref{P0-C-amplification power}), (\ref{SINR1}), (\ref{SINR2}), and (\ref{SINR3}) is difficult to solve.
	To address the CSI uncertainty, we introduce two useful lemmas below to reformulate constraints (\ref{P0-C-amplification power}), (\ref{SINR1}), (\ref{SINR2}), and (\ref{SINR3}), respectively.
	
	\begin{lemma}
		\label{lemma1}
		(General S-Procedure \cite{Zhou2020CSI}) Let $\mathbf{B}_i \in \mathbb{H}^{N}$. Define $\mathbf{x} \in \mathbb{C}^{N \times 1}$ with:
		\begin{align}
			f_i(\mathbf{x})=\mathbf{x}^{\rm H}\mathbf{B}_i\mathbf{x}+2\Re\{\mathbf{b}_i^{\rm H}\mathbf{x}\}+b_i,~i=0,\ldots,P. 
		\end{align}
		The condition $\{f_i(\mathbf{x}) \ge 0\}_{i=1}^P \rightarrow f_0(\mathbf{x}) \ge 0$ holds when for each $i$, there exists $\omega_i \ge 0$ such that
		\begin{align}
			\begin{bmatrix}
				\mathbf{B}_0 & \mathbf{b}_0 \\
				\mathbf{b}_0^{\rm H}  & b_0
			\end{bmatrix}-\sum\nolimits_{i=1}^P \omega_i 
			\begin{bmatrix}
				\mathbf{B}_i & \mathbf{b}_i \\
				\mathbf{b}_i^{\rm H}  & b_i
			\end{bmatrix} \succeq \mathbf{0}. 
		\end{align}	
	\end{lemma}
	
	\begin{lemma}
		\label{lemma2}
		(General sign-definiteness \cite{Zhou2020CSI}) For a given set of matrices $\mathbf{B}=\mathbf{B}^{\mathrm{H}}$, and $\{\mathbf{A}_i,\mathbf{D}_i\}_{i=1}^P$, the following linear matrix inequality (LMI) holds
		\begin{align}
			&\!\!\! \mathbf{B} \succeq \sum\nolimits_{i=1}^P (\mathbf{A}_i^{\rm H}\mathbf{X}_i\mathbf{D}_i+\mathbf{D}_i^{\rm H}\mathbf{X}_i^{\rm H}\mathbf{A}_i),~\forall i, ~\Vert \mathbf{X}_i \Vert_F \le \zeta_i,
		\end{align}
		if and only if for each $i$, there exists a real number $\upsilon_i \ge 0$ such that
		\begin{align}
			&\!\!\!\!\! \begin{bmatrix}
				\mathbf{B}-\sum\nolimits_{i=1}^P \upsilon_i\mathbf{D}_i^{\rm H}\mathbf{D}_i & -\zeta_1\mathbf{A}_1^{\rm H} & \cdots & -\zeta_P\mathbf{A}_P^{\rm H} \\
				-\zeta_1\mathbf{A}_1  & \upsilon_1\mathbf{I} & \cdots & \mathbf{0} \\
				\vdots & \vdots & \ddots & \vdots \\
				-\zeta_P\mathbf{A}_P  & \mathbf{0} & \cdots & -\upsilon_P\mathbf{I}
			\end{bmatrix} \succeq \mathbf{0}.  
		\end{align}
	\end{lemma}
	
	To handle constraint (\ref{SINR1}), we denote $(\mathbf{w}_{c,k}^{(\tau)}, ~\mathbf{u}_c^{(\tau)})$ as the optimal solutions obtained in the $\tau$-th iteration. By replacing $\mathbf{h}_{c,k}$ and $\mathbf{F}_{c,k}$ with $\mathbf{h}_{c,k}=\mathbf{\widehat{h}}_{c,k}+\Delta\mathbf{h}_{c,k}$ and $\mathbf{F}_{c,k}=\mathbf{\widehat{F}}_{c,k}+\Delta\mathbf{F}_{c,k}$, constraint (\ref{SINR1}) is equivalently rewritten as 
	\begin{align}
		\label{eqn16}
		\mathbf{x}_{c,k}^{\rm H} \mathbf{A}_{c,k} \mathbf{x}_{c,k} +& 2\Re\{\mathbf{a}_{c,k}^{\rm H}\mathbf{x}_{c,k}\}+  a_{c,k} \ge \alpha_{c,k}, \nonumber \\
		&\Lambda_{h,c,k}, ~\Lambda_{F,c,k}, ~\forall c, ~\forall k. 
	\end{align}
	The involved derivation is presented in Appendix \ref{Appendix A}.
	
	Based on Lemma \ref{lemma1}, we equivalently recast $\Lambda_{h,c}$ and $\Lambda_{F,c}$ as
	\begin{align}
		\left\{
		\begin{aligned}
			&\mathbf{x}_{c,k}^{\rm H}
			\begin{bmatrix}
				\mathbf{I}_N & \mathbf{0} \\
				\mathbf{0}  & \mathbf{0}
			\end{bmatrix}
			\mathbf{x}_{c,k} - \xi_{h,c,k}^2 \le 0, \\
			&\mathbf{x}_{c,k}^{\rm H}
			\begin{bmatrix}
				\mathbf{0}  & \mathbf{0} \\
				\mathbf{0}  & \mathbf{I}_{MN}
			\end{bmatrix}
			\mathbf{x}_{c,k} - \xi_{F,c,k}^2 \le 0.
		\end{aligned}
		\right.  
	\end{align}
	Then, we transform constraint (\ref{SINR1}) into equivalent LMIs \cite{Hanif2014} by introducing slack variables $\bm{\omega}_h=[\omega_{h,c,1},\ldots,\omega_{h,c,K_c},\omega_{h,\bar{c}, 1},\ldots,\omega_{h,\bar{c},K_{\bar{c}}}]^{\rm T} \ge 0$ and
	$\bm{\omega}_F=[\omega_{F,c,1},\ldots,\omega_{F,c,K_c},\omega_{F,\bar{c}, 1},\ldots,\omega_{F,\bar{c},K_{\bar{c}}}]^{\rm T} \ge 0$, which is given by
	\begin{align}
		\label{SINR1-2}
		& \!\!\!\!\!\! \begin{bmatrix}
			\mathbf{A}_{c,k}+
			\begin{bmatrix}
				\omega_{h,c,k}\mathbf{I}_N &  \mathbf{0}  \\
				\mathbf{0}   &   \omega_{F,c,k}\mathbf{I}_{MN}
			\end{bmatrix}
			& \mathbf{a}_{c,k} \\
			\mathbf{a}_{c,k}^{\rm H}  & A_{c,k}
		\end{bmatrix}  \succeq \mathbf{0}, ~\forall c,~\forall k,  
	\end{align}
	where $A_{c,k}=a_{c,k}-\alpha_{c,k}-\omega_{h,c,k}\xi_{h,c,k}^2-\omega_{F,c,k}\xi_{F,c,k}^2$. 
	Constraint (\ref{P0-C-amplification power}) can be equivalently recast as 
	\begin{align}
		&\!\!\!\! \sum\nolimits_{c}\!\!(\Vert \boldsymbol{\Theta}_c \mathbf{G}\mathbf{W}  \Vert_F^2 \!+\! \Vert \boldsymbol{\Theta}_c \Vert_F ^2 \sigma_1^2) \nonumber \\ 
		&\!\!\!\! = \!\!\!\sum\nolimits_{c} \!(  \Vert \boldsymbol{\Theta}_c \mathbf{G} \sum\nolimits_{\tilde{c}}\sum\nolimits_{k=1}^{K_c}\mathbf{w}_{\tilde{c},k}  \Vert ^2 \!+\! \Vert \boldsymbol{\Theta}_c \Vert_F ^2 \sigma_1^2) \leq P_{\mathrm{RIS}}^{\mathrm{max}}, 
	\end{align}
	where $\tilde{c} \in \{r,t\}$. By substituting $\mathbf{G}=\mathbf{\widehat{G}}+\Delta\mathbf{G}$, we have \cite{Wang2023CSI}
	\begin{align}
		\label{amplification power1}
		&\!\!\!\!\! \sum\nolimits_{c}(\mathbf{y}_c^{\mathrm{H}}\mathbf{B}\mathbf{y}_c + 2\Re\{\mathbf{y}_c^{\mathrm{H}}\mathbf{B}\widehat{\mathbf{y}}_c\} +\widehat{\mathbf{y}}_c^{\mathrm{H}}\mathbf{B}\widehat{\mathbf{y}}_c) + b \le 0, ~\Lambda_{G},
	\end{align}
	where
	\begin{align}
		&\!\!\!\! \mathbf{y}_c=\mathrm{vec}(\boldsymbol{\Theta}_c\widehat{\mathbf{G}}),  ~b=-P_{\mathrm{RIS}}^{\mathrm{max}} + \sum\nolimits_{c}\Vert \boldsymbol{\Theta}_c \Vert_F ^2 \sigma_1^2, \nonumber \\
		&\!\!\!\! \mathbf{B}_{r/t} \!=\! \mathbf{I}_M \!\otimes\! (\sum\nolimits_{\tilde{c}}\!\sum\nolimits_{k=1}^{K_c} \!\!\mathbf{w}_{\tilde{c},k} \mathbf{w}_{\tilde{c},k}^{\mathrm{H}} ), ~\widehat{\mathbf{y}}_c \!=\!\mathrm{vec}(\boldsymbol{\Theta}_c\Delta\mathbf{G}). \nonumber
	\end{align}
	Denote
	\begin{align}
		& \widetilde{\mathbf{y}}=	\begin{bmatrix}
			\widehat{\mathbf{y}}_r^{\mathrm{H}} \! \quad \! 
			\widehat{\mathbf{y}}_t^{\mathrm{H}}
		\end{bmatrix}^{\mathrm{H}},
		\overline{\mathbf{y}}=	\begin{bmatrix}
			\mathbf{y}_r^{\mathrm{H}} \! \quad \!
			\mathbf{y}_t^{\mathrm{H}}
		\end{bmatrix}^{\mathrm{H}}, ~ \mathrm{and} ~ 
		\mathbf{B} \!=\! \begin{bmatrix}
			\mathbf{B}_r \!\!&\!\! \mathbf{0}_{MN} \\
			\mathbf{0}_{MN} \!\!&\!\! \mathbf{B}_t
		\end{bmatrix}, \nonumber
	\end{align}
	then we have the following inequality
	\begin{align}
		\label{amplification power3}
		&\!\!\!\!\! \widetilde{\mathbf{y}}^{\mathrm{H}}\mathbf{B}\widetilde{\mathbf{y}} + 2\Re\{\overline{\mathbf{y}}^{\mathrm{H}}\mathbf{B}\widetilde{\mathbf{y}}\} +\overline{\mathbf{y}}^{\mathrm{H}}\mathbf{B}\overline{\mathbf{y}} + b \le 0, ~\Lambda_{G}.
	\end{align}
	Based on $\mathbf{G}=\mathbf{\widehat{G}}+\Delta\mathbf{G}$ and $\Vert \Delta\mathbf{G} \Vert_F \le \xi_{G}$, we obtain $\Vert \mathrm{vec}(\boldsymbol{\Theta}_c\Delta\mathbf{G})\Vert \le \frac{\xi_{G} \Vert \boldsymbol{\Theta}_c \Vert_F}{\sqrt{M}}$. Thus, we have
	\begin{align}
		\left\{
		\begin{aligned}
			&\widetilde{\mathbf{y}}^{\mathrm{H}}
			\begin{bmatrix}
				\mathbf{I}_{MN} & \mathbf{0} \\
				\mathbf{0}  & \mathbf{0}
			\end{bmatrix}
			\widetilde{\mathbf{y}} - \frac{\xi_{G}^2 \Vert \boldsymbol{\Theta}_r \Vert_F^2}{M} \le 0, \\
			&\widetilde{\mathbf{y}}^{\mathrm{H}}
			\begin{bmatrix}
				\mathbf{0}  & \mathbf{0} \\
				\mathbf{0}  & \mathbf{I}_{MN}
			\end{bmatrix}
			\widetilde{\mathbf{y}} - \frac{\xi_{G}^2 \Vert \boldsymbol{\Theta}_t \Vert_F^2}{M} \le 0.
		\end{aligned}
		\right.  
	\end{align}
	
	According to Lemma \ref{lemma1}, by introducing slack variables $\varsigma_G^r \ge 0$ and $\varsigma_G^t \ge 0$, constraint (\ref{amplification power3}) is transformed into the following LMI constraint:
	\begin{align}
		\label{amplification power2}
		& \!\!\!\!\!\! \begin{bmatrix}
			\begin{bmatrix}
				\varsigma_G^r\mathbf{I}_{MN} &  \mathbf{0}  \\
				\mathbf{0}   &   \varsigma_G^t\mathbf{I}_{MN}
			\end{bmatrix} -\mathbf{B}
			& -\mathbf{B}^{\mathrm{H}} \overline{\mathbf{y}} \\
			-\overline{\mathbf{y}}^{\mathrm{H}} \mathbf{B}  & C_G
		\end{bmatrix}  \succeq \mathbf{0}, ~ \Lambda_{G},  
	\end{align}
	where $C_G = -\overline{\mathbf{y}}^{\mathrm{H}} \mathbf{B} \overline{\mathbf{y}} - b -\frac{\varsigma_G^r \xi_{G}^2\Vert \boldsymbol{\Theta}_r \Vert_F^2}{M}-\frac{\varsigma_G^t \xi_{G}^2 \Vert \boldsymbol{\Theta}_t \Vert_F^2}{M}$.

	To tackle the term $P_{\mathrm{total}} \le \varrho$ in constraint (\ref{P1-P_t}), we have 
	\begin{align}
		&\!\!\!\! \widetilde{P}_{\mathrm{RIS}} \!=\!  \widetilde{\mathbf{y}}^{\mathrm{H}}\mathbf{B}\widetilde{\mathbf{y}} \!+\! 2\Re\{\overline{\mathbf{y}}_c^{\mathrm{H}}\mathbf{B}\widetilde{\mathbf{y}}\} \!+\!\overline{\mathbf{y}}^{\mathrm{H}}\mathbf{B}\overline{\mathbf{y}} \!+\! \sum\nolimits_{c}\!\! \Vert \boldsymbol{\Theta}_c \Vert_F ^2 \sigma_1^2. 
	\end{align}
	Then we have 
	\begin{align}
		\label{P-RIS}
		\widetilde{P}_{\mathrm{total}} =& \xi \sum\nolimits_{c}\sum\nolimits_{k=1}^{K_c} \Vert \mathbf{w}_{c,k} \Vert^2 + \zeta \widetilde{P}_{\mathrm{RIS}} \nonumber \\
		& +  KP_U +  P_{\mathrm{BS}} + 2M(P_{\mathrm{PS}} + P_{\mathrm{PA}}) \le \varrho. 
	\end{align}
	Similar to (\ref{amplification power2}), by introducing auxiliary variables $\vartheta_G^r \ge 0$ and $\vartheta_G^t \ge 0$, constraint (\ref{P-RIS}) can be given by
	\begin{align}
		\label{P-RIS1}
		& \!\!\!\!\!\!  \begin{bmatrix}
			\begin{bmatrix}
				\vartheta_G^r\mathbf{I}_{MN} &  \mathbf{0}  \\
				\mathbf{0}   &   \vartheta_G^t\mathbf{I}_{MN}
			\end{bmatrix} -\mathbf{B}
			& -\mathbf{B}^{\mathrm{H}} \overline{\mathbf{y}} \\
			-\overline{\mathbf{y}}^{\mathrm{H}} \mathbf{B}  & C_P
		\end{bmatrix}  \succeq \mathbf{0}, ~ \Lambda_{G}, ~\forall c,~\forall k,  
	\end{align}
	where $C_P = -\overline{\mathbf{y}}^{\mathrm{H}} \mathbf{B} \overline{\mathbf{y}} - \widehat{\mathrm{P}}_{\mathrm{RIS}} -\frac{\vartheta_G^r \xi_{G}^2 \Vert \boldsymbol{\Theta}_r \Vert_F^2}{M}-\frac{\vartheta_G^t \xi_{G}^2\Vert \boldsymbol{\Theta}_t \Vert_F^2}{M}$, and $\widehat{\mathrm{P}}_{\mathrm{RIS}} = \frac{1}{\zeta}(\xi \sum\nolimits_{c}\sum\nolimits_{k=1}^{K_c} \Vert \mathbf{w}_{c,k} \Vert^2 +  KP_U +  P_{\mathrm{BS}} + M(P_{\mathrm{PS}} + P_{\mathrm{PA}}) - \varrho) + \sum\nolimits_{c}\Vert \boldsymbol{\Theta}_c \Vert_F ^2 \sigma_1^2$.
	
	Then, we tackle the uncertainties in $\{\Delta\mathbf{h}_{c,k},~\Delta\mathbf{F}_{c,k}, ~\Delta\mathbf{f}_{c,k}\}$ of (\ref{SINR2}) and (\ref{SINR3}).
	Based on Lemma \ref{lemma2}, we define slack variables $\bm{\upsilon}_{F}=[\upsilon_{F,c,1},\ldots,\upsilon_{F,c,K_c},\upsilon_{F,\bar{c}, 1},\ldots,\upsilon_{F,\bar{c}, K_{\bar{c}}}]^{\rm T} \ge 0$, $\bm{\upsilon}_{h} =[\upsilon_{h,c,1},\ldots,\upsilon_{h,c,K_c},\upsilon_{h,\bar{c}, 1},\ldots,\upsilon_{h,\bar{c}, K_{\bar{c}}}]^{\rm T} \ge 0$, and $\bm{\upsilon}_{f} =[\upsilon_{f,c,1},\ldots,\upsilon_{f,c,K_c},\upsilon_{f,\bar{c}, 1},\ldots,\upsilon_{f,\bar{c}, K_{\bar{c}}}]^{\rm T} \ge 0$, then the equivalent LMIs of constraints (\ref{SINR2}) and (\ref{SINR3}) are given by
	\begin{subequations}
		\begin{align}
			\label{SINR2-1}
			& \!\!\!\!\!\!\! \begin{bmatrix}
				T_{c,k} \!&\! \mathbf{t}_{c,k} \!&\! \mathbf{0}_{1 \times N} \!&\!\! \mathbf{0}_{1 \times N}  \\
				\mathbf{t}_{c,k}^{\mathrm{H}}  \!&\! \mathbf{I}_{K-1} \!&\!\! \xi_{F,c,k}\mathbf{W}_{c,-k}^{\rm H} \!& \xi_{h,c,k}\mathbf{W}_{c,-k}^{\rm H}  \\
				\mathbf{0}_{N \times 1} \! &\! \xi_{F,c,k}\mathbf{W}_{c,-k} \!&\!\! \upsilon_{F,c,k}\mathbf{I}_N \!&\! \mathbf{0}_{N \times N}  \\
				\mathbf{0}_{N \times 1} \!&\! \xi_{h,c,k}\mathbf{W}_{c,-k}  \!&\!\! \mathbf{0}_{N \times N} \!&\! \upsilon_{h,c,k}\mathbf{I}_N 
			\end{bmatrix}
			\! \succeq \! \mathbf{0},  
		\end{align}
		\begin{align}
			\label{SINR3-1}
			&\!\!\!\!\!\!\! \begin{bmatrix}
				\mathcal{A}_{c,k}-\upsilon_{f,c,k} & \sigma_1 \mathbf{f}_{c,k}\boldsymbol{\Theta}_c & \mathbf{0}_{1 \times M}  \\
				\sigma_1 \boldsymbol{\Theta}_c^{\mathrm{H}}\mathbf{f}_{c,k}^{\mathrm{H}}  & \mathbf{I}_M & \xi_{f,c,k}\sigma_1\boldsymbol{\Theta}_c^{\rm H}   \\
				\mathbf{0}_{M \times 1} & \xi_{f,c,k}\sigma_1\boldsymbol{\Theta}_c & \upsilon_{f,c,k}\mathbf{I}_M 
			\end{bmatrix}
			\succeq \mathbf{0}, ~\forall c,~\forall k, 
		\end{align}
	\end{subequations}
	where $T_{c,k}=\eta_{c,k}-\mathcal{A}_{c,k}-\sigma_2^2-\upsilon_{h,c,k}-\upsilon_{F,c,k}\sum\nolimits_{m=1}^M\beta_m^c$. The detailed transformations are provided in Appendix \ref{Appendix B}.

	After replacing constraints (\ref{P0-C-amplification power}), (\ref{SINR1}), (\ref{SINR2}), and (\ref{SINR3}) with (\ref{amplification power2}), (\ref{SINR1-2}), (\ref{SINR2-1}), and (\ref{SINR3-1}), respectively, we rewrite problem (\ref{P0}) as follows:
	\begin{subequations}
		\label{P3}
		\begin{eqnarray}
			&\!\!\!\!\!\!\!\!\!\!\!\!\!\!\!\!\! \max \limits_{\mathbf{w}_{c,k},\boldsymbol{\Theta}_c, \Delta} 
			\label{P3-function}
			&\!\!\!\!\!\!\!  \psi  \\
			\label{P3-0} 
			&\!\!\!\!\!\!\!\!\!\!\!\!\!\!\!\!\!  \mathrm{s.t.} & \!\!\!\!\!\!\! (\rm{\ref{P0-C-transmit power}}),~(\rm{\ref{P0-C-RIS-coefficients}}), ~(\rm{\ref{P0-C-RIS}}),~(\rm{\ref{P2-1}}), ~(\rm{\ref{P2-r}}), \\
			&&\!\!\!\!\!\!\! (\rm{\ref{SINR0-1}}),~(\rm{\ref{SINR1-2}}), ~(\rm{\ref{amplification power2}}),~(\rm{\ref{P-RIS1}}), ~ (\rm{\ref{SINR2-1}}), ~(\rm{\ref{SINR3-1}}),\\
			\label{auxiliary1}
			&&\!\!\!\!\!\!\! \omega_{h,c,k}, \omega_{F,c,k}, \upsilon_{F,c,k}, \upsilon_{h,c,k}, \upsilon_{f,c,k} \ge 0, \forall c, ~\forall k, \\
			\label{auxiliary}
			&&\!\!\!\!\!\!\! \varsigma_G^r,\varsigma_G^t,\vartheta_G^r, \vartheta_G^t \ge 0,
		\end{eqnarray}
	\end{subequations}
	where $\Delta = \{\psi, \varrho, \mathbf{r}, \bm{\eta}, \bm{\omega}_h, \bm{\omega}_F, \bm{\upsilon}_{F}, \bm{\upsilon}_{h}, \bm{\upsilon}_{f}, \varsigma_G^r,\varsigma_G^t,\vartheta_G^r, \vartheta_G^t, x_{c,k}^{(p)}\}$.
	
	Since problem (\ref{P3}) is still difficult to address directly due to the highly coupled optimization variables $\mathbf{w}_{c,k}$ and $\boldsymbol{\Theta}_c$, we apply an AO algorithm to deal with it in the following.
	
	\vspace{-2mm}
	\subsection{Transmit Beamforming Design}
	\vspace{-1mm}
	For fixed $\boldsymbol{\Theta}_c$, we optimize the transmit beamforming vector $\mathbf{w}_{c,k}$ in this subsection. 
	The corresponding optimization problem is formulated as
	\begin{subequations}
		\label{P3-1}
		\begin{eqnarray}
			&\!\!\!\!\!\!\!\!\!\!\!\!\!\!\!\!\!\!\!\!\!\!\!\! \max \limits_{\mathbf{w}_{c,k}, \Delta} 
			\label{P3-1-function}
			&\!\!\!\!\!  \psi  \\
			\label{P3-1-0} 
			&\!\!\!\!\!\!\!\!\!\!\!\!\!\!\!\!\!\!\!\!\!\!\!\!  \mathrm{s.t.} & \!\!\!\!\! (\rm{\ref{P0-C-transmit power}}), ~(\rm{\ref{P2-1}}), ~(\rm{\ref{P2-r}}),~(\rm{\ref{SINR0-1}}),~(\rm{\ref{SINR1-2}}), ~(\rm{\ref{amplification power2}}), \\
			&&\!\!\!\!\! (\rm{\ref{P-RIS1}}), ~ (\rm{\ref{SINR2-1}}), ~(\rm{\ref{SINR3-1}}),~ (\rm{\ref{auxiliary1}}),~ (\rm{\ref{auxiliary}}).
		\end{eqnarray}
	\end{subequations}
	Problem (\ref{P3-1}) is an semidefinite programming (SDP) problem and can be solved via CVX \cite{Boyd2004CVX}.
	
	\vspace{-3mm}
	\subsection{MF-RIS Coefficients Design}
	\vspace{-1mm}
	For fixed $\mathbf{w}_{c,k}$, we optimize the MF-RIS coefficients $\boldsymbol{\Theta}_c$ in this subsection. 
	The corresponding subproblem is given by
	\begin{subequations}
		\label{P3-2}
		\begin{eqnarray}
			&\!\!\!\!\!\!\!\!\! \max \limits_{\boldsymbol{\Theta}_c, \Delta} 
			\label{P3-2-function}
			&\!\!  \psi  \\
			\label{P3-2-amplitude}
			&\!\!\!\!\!\!\!\!\!  \mathrm{s.t.} & \!\! [\mathbf{u}_c]_m = \sqrt{\beta_m^c}e^{j\theta_m^c}, \forall c,~\forall m,\\
			\label{P3-2-0}
			&&\!\! (\rm{\ref{P0-C-RIS-coefficients}}),~(\rm{\ref{P0-C-RIS}}), ~(\rm{\ref{P2-1}}), ~(\rm{\ref{P2-r}}),~(\rm{\ref{SINR0-1}}), ~(\rm{\ref{amplification power2}}),\\
			&&\!\! (\rm{\ref{P-RIS1}}), ~ (\rm{\ref{SINR2-1}}), ~(\rm{\ref{SINR3-1}}), ~ (\rm{\ref{auxiliary1}}),~ (\rm{\ref{auxiliary}}).
		\end{eqnarray}
	\end{subequations}
	For constraint (\ref{amplification power2}), we have $\frac{\varsigma_G^c \xi_{G}^2 \Vert \boldsymbol{\Theta}_c \Vert_F^2}{M} =\frac{\xi_{G}^2}{M} \varsigma_G^c \mathbf{u}_c^{\mathrm{H}}\mathbf{u}_c$. Then, denote $u_c=\mathbf{u}_c^{\mathrm{H}}\mathbf{u}_c$,
	we apply the convex upper bound approximation to tackle the non-convex term $\varsigma_G^c u_c$ \cite{Tran2012CUB}. Thus, when $s=\varsigma_G^c / u_c $, we have
	\begin{align}
		\label{Trans1}
		G(\varsigma_G^c,u_c) = \frac{s}{2}(u_c)^2 + \frac{(\varsigma_G^c)^2}{2s} \ge \varsigma_G^c u_c, ~s \ge 0.
	\end{align}	
	The transformation of term $\frac{\vartheta_G^r \xi_{G}^2 \Vert \boldsymbol{\Theta}_c \Vert_F^2}{M}$ in constraint (\ref{P-RIS1}) is the same as that of term $\frac{\zeta_G^c \xi_{G}^2 \Vert \boldsymbol{\Theta}_c \Vert_F^2}{M}$ in constraint (\ref{Trans1}). Thus, we recast constraints (\ref{amplification power2}) and (\ref{P-RIS1}) as $(\ref{amplification power2}^{\prime})$ and $(\ref{P-RIS1}^{\prime})$, respectively, which are omitted here for brevity.
	For constraint (\ref{P3-2-amplitude}), we introduce a variable set $\mathbf{b} = \{b_{c,m}, \forall (c,m)\}$, satisfying $b_{c,m} = [\mathbf{u}_c]_m^{\ast}[\mathbf{u}_c]_m$. Then we have $b_{c,m} \le [\mathbf{u}_c]_m^{\ast}[\mathbf{u}_c]_m \le b_{c,m}$. By adopting SCA approximation, the left term is rewritten as $b_{c,m} \le 2\Re\{[\mathbf{u}_c]_m^{\ast}[\mathbf{u}_c^{(\tau)}]_m\} - [\mathbf{u}_c^{(\tau)}]_m^{\ast}[\mathbf{u}_c^{(\tau)}]_m$.
	Then, by introducing a slack variable set $\mathbf{d}=\{d_{c,m},~\widehat{d}_{c,m} ,~ \forall m\}$, problem (\ref{P3-2}) is reformulated as
	\begin{subequations}
		\label{P3-4}
		\begin{eqnarray}
			&\!\!\!\!\!\!\!\!\!\!\!\!\!\!\!\!\! \max \limits_{\boldsymbol{\Theta}_c, \Delta,\mathbf{b}, \mathbf{d}} 
			\label{P3-4-function}
			& \!\!\!\!\!  \psi-\lambda^{(\tau)} D  \\
			\label{P3-4-0} 
			&\!\!\!\!\!\!\!\!\!\!\!\!\!\!\!\!\!  \mathrm{s.t.} & \!\!\!\!\!\!\!\!\!\!
			[\mathbf{u}_c]_m^{\ast}[\mathbf{u}_c]_m \le b_{c,m} + d_{c,m}, ~b_{c,m} \in [0,\beta_{\mathrm{max}}],  \\
			\label{P3-4-1} 
			&&\!\!\!\!\!\!\!\!\!\! 2\Re\{[\mathbf{u}_c]_m^{\ast}[\mathbf{u}_c^{(\tau)}]_m\} \!\!-\!\! [\mathbf{u}_c^{(\tau)}]_m^{\ast}[\mathbf{u}_c^{(\tau)}]_m \!\ge\! b_{c,m} \!-\! \widehat{d}_{c,m},\\
			&&\!\!\!\!\!\!\!\!\!\! \sum\nolimits_{c} b_{c,m} \le  \beta_{\mathrm{max}},~ (\rm{\ref{P0-C-RIS-coefficients}}), ~(\rm{\ref{P0-C-RIS}}),~(\rm{\ref{P2-1}}), ~(\rm{\ref{P2-r}}), \\
			&&\!\!\!\!\!\!\!\!\!\! (\rm{\ref{SINR0-1}}),~(\rm{\ref{amplification power2}^{\prime}}),~(\rm{\ref{P-RIS1}^{\prime}}),~ (\rm{\ref{SINR2-1}}), ~(\rm{\ref{SINR3-1}}), ~ (\rm{\ref{auxiliary1}}), ~ (\rm{\ref{auxiliary}}),
		\end{eqnarray}
	\end{subequations}
	where $\lambda$ is the scaling factor to manage the feasibility of constraints $(\mathrm{\ref{P3-4-0}})$ and  $(\mathrm{\ref{P3-4-1}})$, and $D = \sum_{c}\sum_{m}(d_{c,m}+\widehat{d}_{c,m})$ is the penalty term. 
	Problem (\ref{P3-4}) is a SDP problem and can be solved via CVX.
	
	\begin{algorithm}[t]  
		\caption{PCCP-Based Algorithm for Solving (\ref{P3-4})}
		\label{Algorithm}
		\renewcommand{\algorithmicrequire}{\textbf{Initialize}}
		\renewcommand{\algorithmicensure}{\textbf{Output}}
		\begin{algorithmic}[1]
			\STATE \textbf{Initialize} the iteration $\tau = 0$ and $\mathbf{u}_c^{(0)}$, the scaling factor $\varepsilon > 1$, the maximum value $\lambda_{\mathrm{max}}$, the allowance tolerances $\epsilon_1$ and $\epsilon_2$, and the maximum iterations $T_{\mathrm{max}}$.
			\REPEAT 
			\IF{$\tau < T_{\mathrm{max}}$} 
			\STATE Update $\mathbf{u}_c^{(\tau+1)}$ by solving (\ref{P3-4});
			\STATE Update $\lambda^{(\tau+1)} = \rm{min}\{\varepsilon \lambda^{(\tau)},  \lambda_{\mathrm{max}}\}$;
			\STATE Update $\tau = \tau +1$;
			\ELSE
			\STATE Reinitialize a new  $\mathbf{u}_c^{(0)}$, and set $\varepsilon > 1$ and $\tau =0$.
			\ENDIF
			\UNTIL $\Vert \mathbf{u}_c^{(\tau)} - \mathbf{u}_c^{(\tau-1)} \Vert_1 \le \epsilon_1 $ and $D \le \epsilon_2$;
			\STATE \textbf{Output} $\mathbf{u}_c^{\ast} = \mathbf{u}_c^{(\tau)}$.
		\end{algorithmic}
	\end{algorithm}	
	
	The detailed procedure of solving problem (\ref{P3-4}) is given in Algorithm \ref{Algorithm}. In the proposed PCCP-based algorithm, constraint (\ref{P3-2-amplitude}) is guaranteed by $D \le \epsilon_2$ when $\epsilon_2$ is sufficiently small. Moreover, the maximum value $\lambda_{\mathrm{max}}$ is exploited to provide an upper bound. In particular, a feasible solution satisfying $D \le \epsilon_2$ may not be found when the iteration converges under increasing $\lambda^{(\tau)}$. In addition, if the maximum number of iterations $T_{\mathrm{max}}$ is reached, we find a new initial point to restart the iteration.
	Therefore, the original problem (\ref{P0}) is tackled by solving subproblems (\ref{P3-1}) and (\ref{P3-4}) alternately, where Algorithm \ref{Algorithm} is applied to handle $\boldsymbol{\Theta}_c$.
	
	\section{Robust Beamforming for Statistical CSI Error} \label{statistical}
	Generally, channel estimation errors follow the Gaussian distribution in practice \cite{Wang2020CSI}, which is unbounded. Consequently, the aforementioned bounded channel model may not accurately express the real-world channel errors. In this section, we study the effect of estimation errors on system performance based on the statistical CSI error model.
	
	\subsection{Problem Formulation}
	For the statistical CSI error model, we consider the data rate outage constraints at users. 
	To derive an analytical expression for the average MF-RIS amplification power, we provide Lemma \ref{lemma5} as follows.
	\begin{lemma}
		\label{lemma5} 
		For given matrices $\mathbf{X} \in \mathbb{C}^{N \times N}$ and $\mathbf{G}=\widehat{\mathbf{G}}+\mathbf{A}^{\frac{1}{2}}\mathbf{G}^{w}\mathbf{B}^{\frac{T}{2}} \in \mathbb{C}^{M \times N}$ with mean $\widehat{\mathbf{G}}$ and covariance $\mathbf{A} \otimes \mathbf{B}$, where $\mathbf{A} = \mathbf{A}^{\frac{1}{2}}\mathbf{A}^{\frac{T}{2}}$, $\mathbf{B} = \mathbf{B}^{\frac{1}{2}}\mathbf{B}^{\frac{T}{2}}$, and $\mathbf{G}^{w}$ is a complex Gaussian random matrix with independent and identically distributed (i.i.d.) entries of zero mean and unit variance \cite{Zhang2018CSI}. We have
		\begin{align}
			\mathbb{E}_{\mathbf{G}}[\mathbf{G}\mathbf{X}\mathbf{G}^{\mathrm{H}}] = \widehat{\mathbf{G}}\mathbf{X}\widehat{\mathbf{G}}^{\mathrm{H}} + \mathrm{Tr}(\mathbf{X}{\mathbf{B}}^{\mathrm{T}})\mathbf{A}.
		\end{align}
	\end{lemma}
	
	Please refer to \cite{Zhang2018CSI} for the proof of Lemma \ref{lemma5}.
	
	We assume $\mathbf{\Sigma}_{G} = \varpi_{G}^2 \mathbf{I}_{MN}$. Then, we have $\mathrm{vec}(\Delta\mathbf{G}) = \varpi_{G} \mathbf{i}_{G}$, where $\mathbf{i}_{G} \sim \mathcal{CN}(\mathbf{0},\mathbf{I}_{MN})$.
	Let $\widetilde{\mathbf{W}} = \mathbf{W} \mathbf{W}^{\mathrm{H}} \in \mathbb{C}^{N \times N}$ and $\mathbf{W}_{c,k} = \mathbf{w}_{c,k} \mathbf{w}_{c,k}^{\mathrm{H}} \in \mathbb{C}^{N \times N}$. Based on $\mathbf{G}=\widehat{\mathbf{G}}+\Delta\mathbf{G}$, we have $\mathbf{G}=\widehat{\mathbf{G}}+\varpi_G^{\frac{1}{2}}\mathbf{I}_M \mathbf{U} \varpi_G^{\frac{1}{2}}\mathbf{I}_N$, where $\mathrm{vec}(\mathbf{U}) \sim \mathcal{CN}(\mathbf{0},\mathbf{I}_{M} \otimes \mathbf{I}_{N}) $ \cite{Zhou2024CSI}. Then, let $\mathbf{v}_c = [\mathbf{u}_c;1]$ and $\widehat{\mathbf{H}}_{c,k} = [\mathbf{F}_{c,k}; \mathbf{h}_{c,k}]$. We define $\mathbf{V}_c = \mathbf{v}_c\mathbf{v}_c^{\mathrm{H}} \in \mathbb{C}^{(M+1) \times (M+1)}$, satisfying $\mathbf{V}_c \succeq \mathbf{0}$ and $\mathrm{Rank}(\mathbf{V}_c) =1$, Thus, we have  $\boldsymbol{\Phi}_c = [\mathbf{V}_c]_{1:M,1:M}$.
	According to Lemma \ref{lemma5}, the first term in constraint (\ref{P0-C-RIS-coefficients}) is recast as:
	\begin{align}
		\label{P-RIS-2}
		\widehat{P}_{\mathrm{RIS}} & = \mathbb{E}_{\mathbf{G}}\{ \sum\nolimits_{c}\Vert \boldsymbol{\Theta}_c \mathbf{G}\mathbf{W}  \Vert_F^2 + \sum\nolimits_{c}\Vert \boldsymbol{\Theta}_c \Vert_F ^2 \sigma_1^2 \} \nonumber \\
		& = \mathrm{Tr}\{\sum\nolimits_{c}\boldsymbol{\Theta}_c  \mathbb{E} \{ \mathbf{G} \mathbf{W} \mathbf{W}^{\mathrm{H}}\mathbf{G}^{\mathrm{H}}\}\boldsymbol{\Theta}_c^{\mathrm{H}} \} + \sum\nolimits_{c}\Vert \boldsymbol{\Theta}_c \Vert_F ^2 \sigma_1^2  \nonumber \\
		& = \mathrm{Tr}\{ \boldsymbol{\Psi}\mathbf{Q}\} + \mathrm{Tr}(\boldsymbol{\Psi}) \sigma_1^2 \le P_{\mathrm{RIS}}^{\mathrm{max}}, 
	\end{align}
	where $\mathbf{Q} =  \widehat{\mathbf{G}}\widetilde{\mathbf{W}}\widehat{\mathbf{G}}^{\mathrm{H}} + \mathrm{Tr}\{\widetilde{\mathbf{W}} \varpi_G\mathbf{I}_N\} \varpi_G\mathbf{I}_M$ with $\boldsymbol{\Psi} = \sum\nolimits_{c}\boldsymbol{\Phi}_c$.
	Then, the corresponding EE maximization problem is given by
	\begin{subequations}
		\label{P4}
		\begin{eqnarray}
			&\!\!\!\!\!\!\!\!\!\!\!\!\!\!\!\! \max \limits_{\mathbf{w}_{c,k},\boldsymbol{\Theta}_c, \mathbf{r}, \psi, \varrho} 
			\label{P4-function}
			&\!\!\!\!\!\!\!\!\!\!\!  \psi  \\
			\label{P4-R} 
			&\!\!\!\!\!\!\!\!\!\!\!\!\!\!\!\!  \mathrm{s.t.} &\!\!\!\!\!\!\!\!\!\!\! \mathrm{Pr} \{R_{c,k} \ge r_{c,k} \} \ge 1-\rho_{c,k}, ~(\ref{statistical CSI}),~\forall c,~\forall k, \\
			\label{P4-P} 
			&&\!\!\!\!\!\!\!\!\!\!\! P_{\mathrm{total}} \le \varrho,~(\rm{\ref{P-RIS-2}}), ~(\ref{statistical CSI}),~\forall c,~\forall k,\\
			&&\!\!\!\!\!\!\!\!\!\!\! (\rm{\ref{P0-C-transmit power}}),~(\rm{\ref{P0-C-RIS-coefficients}}),~(\rm{\ref{P0-C-RIS}}),~(\rm{\ref{P2-1}}),~(\rm{\ref{P2-r}}), ~(\ref{statistical CSI}),
		\end{eqnarray}
	\end{subequations}
	where $\bm{\rho} = \{\rho_{c,1},\rho_{c,2},\ldots,\rho_{c,K_c},\rho_{\bar{c},1},\rho_{\bar{c},1},\ldots,\rho_{\bar{c},K_{\bar{c}}}\}$ denotes the maximum data rate outage probabilities. Similar to problem (\ref{P0}), problem (\ref{P4}) is a non-convex optimization problem. Furthermore, considering that the data rate outage probability constraints do not have closed-form expressions, problem (\ref{P4}) is computationally intractable. Thus, we introduce the Bernstein-Type Inequality, and apply the SROCR and SDR techniques to optimize $\mathbf{w}_{c,k}$ and $\boldsymbol{\Theta}_c$ alternately in the subsequent analysis.

	\subsection{Problem Transformation}
	First, the rate outage probability of user $U_{c,k}$ is transformed as follows:
	
	\begin{align}
		\label{Rate}
		&\!\!\!\! \mathrm{Pr}\left\{\log_{2}\left(1 \!+ \! \frac{\vert \bar{\mathbf{h}}_{c,k}\mathbf{w}_{c,k} \vert^2}
		{\Vert \bar{\mathbf{h}}_{c,k}\mathbf{w}_{c,-k} \Vert_2^2 \!+\! \Vert \mathbf{f}_{c,k}\mathbf{\Theta}_c \Vert_2^2 \sigma_1^2 \!+\! \sigma_2^2}\right) \! \geq \! r_{c,k}\right\} \nonumber \\
		=& \mathrm{Pr}\{\left(\mathbf{h}_{c,k}+\mathbf{u}_{c,k}^{\mathrm{H}}\mathbf{F}_{c,k}\right)\mathbf{C}_{c,k}\left(\mathbf{h}_{c,k}^{\mathrm{H}}+\mathbf{F}_{c,k}^{\mathrm{H}}\mathbf{u}_{c,k}\right) \nonumber \\
		& -\sigma_{1}^{2}(\mathbf{f}_{c,k}\boldsymbol{\Theta}_c)(\boldsymbol{\Theta}_c^{\mathrm{H}}\mathbf{f}_{c,k}^{\mathrm{H}})-\sigma_{2}^{2} \geq 0 \}, ~\forall c,~\forall k,
	\end{align}
	where $\mathbf{C}_{c,k} = \mathbf{w}_{c,k}\mathbf{w}_{c,k}^{\mathrm{H}} / (2^{r_{c,k}} - 1) - \mathbf{W}_{c,-k}\mathbf{W}_{c,-k}^{\mathrm{H}} \in \mathbb{C}^{N \times N} $.
	
	For the sake of facilitating derivations, we assume that $\mathbf{\Sigma}_{h,c,k} = \varpi_{h,c,k}^2 \mathbf{I}_N$, $\mathbf{\Sigma}_{F,c,k} = \varpi_{F,c,k}^2 \mathbf{I}_{MN}$, and $\mathbf{\Sigma}_{f,c,k} = \varpi_{f,c,k}^2 \mathbf{I}_M$. Then, we have $\Delta\mathbf{h}_{c,k} = \varpi_{h,c,k} \mathbf{i}_{h,c,k}$, $\mathrm{vec}(\Delta\mathbf{F}_{F,c,k}) = \varpi_{F,c,k} \mathbf{i}_{F,c,k}$, and $\Delta\mathbf{f}_{c,k} = \varpi_{f,c,k} \mathbf{i}_{f,c,k}$, where $\mathbf{i}_{h,c,k} \sim \mathcal{CN}(\mathbf{0},\mathbf{I}_N)$, $\mathbf{i}_{F,c,k} \sim \mathcal{CN}(\mathbf{0},\mathbf{I}_{MN})$, and $\mathbf{i}_{f,c,k} \sim \mathcal{CN}(\mathbf{0},\mathbf{I}_M)$.
	To address problem (\ref{P4}), we approximate it based on Bernstein-Type Inequality in the subsequent analysis.
	
	\begin{lemma}
		\label{lemma3}
		(Bernstein-Type Inequality \cite{Wang2014CSI}) Define $\mathbf{x} \in \mathbb{C}^{N \times 1} \sim \mathcal{CN}(\mathbf{0},\mathbf{I}_N)$, $\mathbf{E} \in \mathbb{H}^{N}$, $\mathbf{e} \in \mathbb{C}^{N \times 1}$, and $e \in \mathbb{R}$, we consider $f(\mathbf{x}) = \mathbf{x}^{\mathbf{H}} \mathbf{E} \mathbf{x} +2\Re\{\mathbf{e}^{\mathbf{H}} \mathbf{x}\} +e$. Then, for any $\rho \in [0,1]$, the following approximation holds: 
		\begin{align}
			& \mathrm{Pr}\{ \mathbf{x}^{\mathbf{H}} \mathbf{E} \mathbf{x} +2\Re\{\mathbf{e}^{\mathbf{H}} \mathbf{x}\} +e \ge 0\}  \ge 1-\rho  \nonumber \\
			\Rightarrow & \mathrm{Tr}\{\mathbf{E}\} - \sqrt{2\ln(1 / \rho)} \sqrt{\Vert \mathbf{E} \Vert_F^2 +2\Vert \mathbf{e} \Vert_2^2} \nonumber \\
			& + \ln(\rho)\lambda_{\mathrm{max}}^{+}(-\mathbf{E}) +e \ge 0, \nonumber \\
			\Rightarrow & \left\{ \begin{array}{rcl}
				&\mathrm{Tr}\{\mathbf{E}\} - \sqrt{2\ln(1 / \rho)}x +\ln(\rho)y +e \ge 0 \\
				&\sqrt{\Vert \mathbf{E} \Vert_F^2 +2\Vert \mathbf{e} \Vert_2^2} \le x \\
				&y\mathbf{I}_N +\mathbf{E} \succeq \mathbf{0}, y \ge 0,
			\end{array}
			\right\} \nonumber
		\end{align}
		where $\lambda_{\mathrm{max}}^{+}(-\mathbf{E}) = \mathrm{max}(\lambda_{\mathrm{max}}(-\mathbf{E}),0)$. Here $x$ and $y$ are slack variables. Please refer to \cite{Wang2014CSI} for the proof of Lemma \ref{lemma3}.
	\end{lemma}

	By defining ${e}_{c,k} = (\widehat{\mathbf{h}}_{c,k}+\mathbf{u}_c^{\mathrm{H}}\widehat{\mathbf{F}}_{c,k})\mathbf{C}_{c,k}(\widehat{\mathbf{h}}_{c,k}^{\mathrm{H}}+\widehat{\mathbf{F}}_{c,k}^{\mathrm{H}}\mathbf{u}_) - \widehat{\mathbf{f}}_{c,k}\boldsymbol{\Phi}_c\widehat{\mathbf{f}}_{c,k}^{\mathrm{H}} - \sigma_2^2 $, where $\boldsymbol{\Phi}_c = \boldsymbol{\Theta}_c\boldsymbol{\Theta}_c^{\mathrm{H}}$, the rate outage constraint (\ref{Rate}) is equivalent to
	\begin{align}
		&& \!\!\!\!\!\!\!\!\!\! \mathrm{Pr}\left\{\widetilde{\mathbf{i}}_{c,k}^{\mathrm{H}}{\mathbf{E}}_{c,k}\widetilde{\mathbf{i}}_{c,k}+2\Re\left\{{\mathbf{e}}_{c,k}^{\mathrm{H}}\widetilde{\mathbf{i}}_{c,k}\right\}+{e}_{c,k} \geq 0 \right\} \! \geq \!  1 \!- \!\rho_{c,k}.
	\end{align}
	The specific derivation is presented in Appendix \ref{Appendix C}.
	
	Based on Lemma \ref{lemma3}, by introducing auxiliary variables $\widetilde{\mathbf{x}}=[\widetilde{x}_{c,1},\ldots,\widetilde{x}_{c,K_c},\widetilde{x}_{\bar{c},1},,\ldots,\widetilde{x}_{\bar{c},K_{\bar{c}}}]^{\mathrm{T}}$ and $\widetilde{\mathbf{y}}=[\widetilde{y}_{c,1},\ldots,\widetilde{y}_{c,K_c},\widetilde{y}_{\bar{c},1},,\ldots,\widetilde{y}_{\bar{c},K_{\bar{c}}}]^{\mathrm{T}}$, we approximate the data rate outage constraint of user $U_{c,k}$ as follows:
	\begin{subequations}
		\begin{align}
				&\!\!\!\!\! \mathrm{Tr}\{\mathbf{E}_{c,k}\} \!-\! \sqrt{2\ln(1 / \rho_{c,k})}\widetilde{x}_{c,k} \!+\! \ln(\rho_{c,k})\widetilde{y}_{c,k} \!+\! {e}_{c,k} \ge 0,  \\ 
				&\!\!\!\!\! \sqrt{\Vert \mathbf{E}_{c,k} \Vert_F^2 \!+\! 2\Vert \mathbf{e}_{c,k} \Vert_2^2} \! \le \! \widetilde{x}_{c,k},~\forall c,~\forall k, \\
				&\!\!\!\!\! \widetilde{y}_{c,k}\mathbf{I}_{MN+M+N} \!+\! \mathbf{E}_{c,k} \! \succeq \! \mathbf{0}, ~\widetilde{y}_{c,k} \ge 0,~\forall c,~\forall k. 
		\end{align}
	\end{subequations}
	After some transformation, we have (\ref{E-trans}) at the bottom of the next page \cite{Zhou2020CSI}.
	\begin{figure*}[b]
		\hrulefill  
		\vspace{1mm}
		\normalsize
		\begin{subequations}
			\label{E-trans}
			\begin{align}
				& \mathrm{Tr}\left\{{\mathbf{E}}_{c,k}\right\}=\mathrm{Tr}\Bigl\{\left[\begin{array}{c}\varpi_{h,c,k}\mathbf{C}_{c,k}^a \varpi_{F,c,k}(\mathbf{C}_{c,k}^a \otimes \mathbf{u}_c^{\ast})\end{array}\right] 
				\cdot \left[\begin{array}{cc}\varpi_{h,c,k}\mathbf{C}_{c,k}^b \varpi_{F,c,k}(\mathbf{C}_{c,k}^b\otimes\mathbf{u}_c^\mathrm{T})\end{array}\right]\Bigr\} -\sigma_1^2\varpi_{f,c,k}^2\mathrm{Tr}\{\boldsymbol{\Phi}_{c}\}  \nonumber \\
				& =\! \left(\varpi_{h,c,k}^{2}\!+\! \varpi_{F,c,k}^{2} \sum\nolimits_{m} \!\! \beta_m^c\right)\! \mathrm{Tr}\left\{\mathbf{C}_{c,k}\right\} \!-\! \sigma_1^2\varpi_{f,c,k}^2\mathrm{Tr}\{\boldsymbol{\Phi}_{c}\}, ~\forall c,~k, \\
				&\Vert {\mathbf{E}}_{c,k}\Vert _{F}^{2} = \widehat{E}_{c,k} =(\varpi_{h,c,k}^{2}+\varpi_{F,c,k}^{2}\sum\nolimits_{m}  \beta_m^c)^{2}\Vert\mathbf{C}_{c,k}\Vert_{F}^{2} - \sigma_1^4\varpi_{f,c,k}^4 \Vert \boldsymbol{\Phi}_{c} \Vert_F^2, ~\forall c,~k,\\
				& \Vert {\mathbf{e}}_{c,k}\Vert ^{2}=\widehat{e}_{c,k} = -\sigma_1^4\varpi_{f,c,k}^2 \Vert \widehat{\mathbf{f}}_{c,k}\boldsymbol{\Phi}_c \Vert_F^2 +\!(\varpi_{h,c,k}^{2}\! +\! \varpi_{F,c,k}^{2}\sum\nolimits_{m}\! \beta_m^c)\Vert (\widehat{\mathbf{h}}_{c,k} \!+\! \mathbf{u}^{\mathrm{H}}\widehat{\mathbf{F}}_{c,k})\mathbf{C}_{c,k}\Vert _{2}^{2},~\forall c,~k, \\
				&\widetilde{y}_{c,k}\mathbf{I}_{MN+M+N}\!+\!{\mathbf{E}}_{c,k} \!\succeq\! \mathbf{0} \Longrightarrow \widetilde{y}_{k}\mathbf{I}_{M+N}\!+\!\mathbf{B}_{c,k} \!\succeq\! \mathbf{0}, \ \mathrm{with} \ \mathbf{B}_{c,k}\!=\!
				\begin{bmatrix} (\varpi_{h,c,k}^{2}\!+\!\varpi_{F,c,k}^{2}\sum\nolimits_{m}\! \beta_m^c)\mathbf{C}_{c,k} \!\!\!&\!\!\! \mathbf{0}_{N \times M} \\
					\mathbf{0}_{M \times N} \!\!\!&\!\!\!	-\sigma_1^2\varpi_{f,c,k}^2 \boldsymbol{\Phi}_{c} 
				\end{bmatrix},
			\end{align}
		\end{subequations}
	\end{figure*}
	where $\mathbf{C}_{c,k} = \mathbf{C}_{c,k}^a\mathbf{C}_{c,k}^b$ with $\mathbf{C}_{c,k}^a = [\sqrt{1/(1-2^{r_{c,k}})}\mathbf{w}_{c,k} \quad {j}\mathbf{w}_{c,-k}]$ and $\mathbf{C}_{c,k}^b = [\sqrt{1/(1-2^{r_{c,k}})}\mathbf{w}_{c,k}^{\mathrm{H}} \quad {j}\mathbf{w}_{c,-k}^{\mathrm{H}}]^{\mathrm{T}}$.
	
	For constraint (\ref{P4-P}), we have 
	\begin{align}
		\label{P-RIS-1}
		P_{\mathrm{total}} =& \xi \sum\nolimits_{c}\sum\nolimits_{k=1}^{K_c} \mathrm{Tr}(\mathbf{W}_{c,k}) + \zeta \widehat{P}_{\mathrm{RIS}} \nonumber \\
		& +  KP_U +  P_{\mathrm{BS}} + 2M(P_{\mathrm{PS}} + P_{\mathrm{PA}}) \le \varrho. 
	\end{align}
	Finally, problem (\ref{P4}) is reformulated as 
	\begin{subequations}
		\label{P5}
		\begin{eqnarray}
			& \!\!\!\!\!\!\!\!\!\!\! \max \limits_{\mathbf{w}_{c,k},\boldsymbol{\Theta}_{c}, \Omega} 
			\label{P5-function}
			&\!\!\!\!\!\!\!\!\!  \psi  \\
			\label{P5-1-1}
			&\!\!\!\!\!\!\!\!\!\!\!  \mathrm{s.t.} &\!\!\!\!\!\!\!\!\!\!\!\! \sqrt{ \widehat{E}_{c,k} +2 \widehat{e}_{c,k}} \le \widetilde{x}_{c,k}, ~\forall c,~\forall k,\\
			\label{P5-1-2}
			&&\!\!\!\!\!\!\!\!\!\!\!\! \left(\varpi_{h,c,k}^{2}\!+\!\varpi_{F,c,k}^{2}\!\! \sum\nolimits_{m}\!\! \beta_m^c\right) \!\mathrm{Tr}\! \left\{\mathbf{C}_{c,k}\right\} \!-\! \sigma_1^2\varpi_{f,c,k}^2 \!\mathrm{Tr}\! \{\boldsymbol{\Phi}_{c}\} \nonumber \\
			&&\!\!\!\!\!\!\!\!\!\!\!\! - \sqrt{2\ln(1 / \rho_{c,k})}\widetilde{x}_{c,k} +\ln(\rho_{c,k})\widetilde{y}_{c,k} +{e}_{c,k} \ge 0,\\
			\label{P5-1-3}
			&&\!\!\!\!\!\!\!\!\!\!\!\! \widetilde{y}_{k}\mathbf{I}_{M+N}+\mathbf{B}_{c,k} \succeq \mathbf{0},~\widetilde{y}_{k} \ge 0, ~\forall c,~\forall k, \\
			&&\!\!\!\!\!\!\!\!\!\!\!\! (\rm{\ref{P0-C-transmit power}}),~(\rm{\ref{P0-C-RIS-coefficients}}), ~(\rm{\ref{P0-C-RIS}}), ~(\rm{\ref{P2-1}}), ~(\rm{\ref{P2-r}}),  ~(\rm{\ref{P-RIS-2}}), ~(\rm{\ref{P-RIS-1}}), 
		\end{eqnarray}
	\end{subequations}
	where $\Omega = \{\mathbf{r}, \psi, \varrho, \widetilde{\mathbf{x}}, \widetilde{\mathbf{y}}\}$.
	To address the coupled optimization variables, we apply the AO framework to tackle problem (\ref{P5}). The resulting sub-problems are optimized by using SDR method in an alternating manner.
	
	\subsection{Transmit Beamforming Design}
	With fixed $\boldsymbol{\Theta}_{c}$, we obtain $\mathbf{W}_{c,k}$ in this subsection.
	Let $\mathbf{C}_{c,k} = \mathbf{W}_{c,k} / (2^{r_{c,k}} - 1) - (\sum\nolimits_{i=1,i \neq k}^{K_c} \mathbf{W}_{c,i} + \sum\nolimits_{i=1}^{K_{\bar{c}}} \mathbf{W}_{\bar{c},i})$. 
	For constraint (\ref{P5-1-1}), we have $\widehat{E}_{c,k} +2\widehat{e}_{c,k} \le \widetilde{x}_{c,k}^2$.
	By adopting the first Taylor expansion, we have $\widetilde{x}_{c,k}^2 \ge \widehat{x}_{c,k} = 2\Re\{(\widetilde{x}_{c,k}^{(\tau),\ast})(\widetilde{x}_{c,k})\}-(\widetilde{x}_{c,k}^{(\tau),\ast})(\widetilde{x}_{c,k}^{(\tau)})$. Then, we obtain the following inequality: $\widehat{E}_{c,k} +2\widehat{e}_{c,k} \le \widehat{x}_{c,k}$.
	The resulting transmit beamforming vector subproblem is given by
	\begin{subequations}
		\label{P5-1}
		\begin{eqnarray}
			& \!\!\!\!\!\!\!\! \max \limits_{\mathbf{W}_{c,k},\Omega} 
			\label{P5-1-function}
			&\!\!\!\!\!\!\!  \psi  \\
			\label{P5-1-E}
			&\!\!\!\!\!\!\!\!  \mathrm{s.t.} &\!\!\!\!\!\!\! \widehat{E}_{c,k}  +2 \widehat{e}_{c,k} \le \widehat{x}_{c,k}, ~\forall c,~\forall k,\\
			&&\!\!\!\!\!\!\! \left(\!\varpi_{h,c,k}^{2}\!+\!\varpi_{F,c,k}^{2}\!\! \sum\nolimits_{m}\! \beta_m\!\right)\!\mathrm{Tr}\left\{\mathbf{C}_{c,k}\right\} \!-\!\sigma_1^2\varpi_{f,c,k}^2\mathrm{Tr}\{\boldsymbol{\Phi}_{c}\} \nonumber \\
			&&\!\!\!\!\!\!\! - \!\sqrt{2\ln(1 / \rho_{c,k})}\widetilde{x}_{c,k} \!+\! \ln(\rho_{c,k})\widetilde{y}_{c,k} \!+\! {e}_{c,k} \!\ge\! 0,\\
			&&\!\!\!\!\!\!\! \widetilde{y}_{k}\mathbf{I}_{M+N}+\mathbf{B}_{c,k} \succeq \mathbf{0},~\widetilde{y}_{k} \ge 0, ~\forall c,\forall k,\\
			\label{P5-1-W_k}
			&&\!\!\!\!\!\!\! \sum\nolimits_{c}\!\sum\nolimits_{k=1}^{K_c} \!\mathrm{Tr}(\mathbf{W}_{c,k}) \!\le\! P_{\mathrm{BS}}^{\mathrm{max}}, ~\mathbf{W}_{c,k} \!\succeq\! \mathbf{0}, ~\forall c, \forall k,\\
			&&\!\!\!\!\!\!\! \mathrm{Rank}({\mathbf{W}_{c,k}}) = 1,~(\rm{\ref{P2-1}}), ~(\rm{\ref{P2-r}}), (\rm{\ref{P-RIS-2}}), (\rm{\ref{P-RIS-1}}).  
		\end{eqnarray}
	\end{subequations}
	
	For the rank-one constraint $\mathrm{Rank}({\mathbf{W}_{c,k}}) = 1$, we exploit the SROCR approach to relax it gradually. Denote $w_{c,k}^{(\tau-1)} \in (0,1]$ as the trace ratio parameter of $\mathbf{W}_{c,k}$ in the $(\tau-1)$-th iteration, then we have \cite{Lyu2023-ARIS}
	\begin{align}
		\label{Rank-W}
		(\mathbf{w}_{c,k}^{(\tau-1)})^{\mathrm{H}}\mathbf{W}_{c,k}^{(\tau)}\mathbf{w}_{c,k}^{(\tau-1)} \ge w_{c,k}^{(\tau-1)} \mathrm{Tr}(\mathbf{W}_{c,k}^{(\tau)}), ~\forall c,~\forall k,
	\end{align}
	where $\mathbf{w}_{c,k}^{(\tau-1)}$ is the eigenvector corresponding to the largest eigenvalue of $\mathbf{W}_{c,k}^{(\tau-1)}$, and $\mathbf{W}_{c,k}^{(\tau-1)}$ is the solution obtained in the $(\tau-1)$-th iteration.
	
	Then, problem (\ref{P5-1}) is reformulated as
	\begin{subequations}
		\label{P5-4}
		\begin{eqnarray}
			& \!\!\!\!\! \max \limits_{\mathbf{W}_{c,k},\Omega} 
			\label{P5-4-function}
			&\!\!\!  \psi  \\
			\label{P5-4-E}
			&\!\!\!\!\!  \mathrm{s.t.} &\!\!\! \widehat{E}_{c,k}  +2 \widehat{e}_{c,k} \le \widehat{x}_{c,k}, ~\forall c,~\forall k,\\
			\label{P5-4-1}
			&&\!\!\! \left(\varpi_{h,c,k}^{2}\!+\!\varpi_{F,c,k}^{2} \!\!\sum\nolimits_{m}\! \beta_m^c\right)\!\mathrm{Tr}\left\{\mathbf{C}_{c,k}\right\} \!-\!\sigma_1^2\varpi_{f,c,k}^2\mathrm{Tr}\{\boldsymbol{\Phi}_{c}\} \nonumber \\
			&&\!\!\! - \sqrt{2\ln(1 / \rho_{c,k})}\widetilde{x}_{c,k} +\ln(\rho_{c,k})\widetilde{y}_{c,k} +{e}_{c,k} \ge 0, \\
			\label{P5-4-2}
			&&\!\!\! \widetilde{y}_{c,k}\mathbf{I}_{M+N} + \mathbf{B}_{c,k} \succeq \mathbf{0},~\forall c,~k, ~\widetilde{y}_{k} \ge 0,~\forall c,~\forall k,\\
			&&\!\!\! (\rm{\ref{P2-1}}), ~(\rm{\ref{P2-r}}), ~(\rm{\ref{P-RIS-2}}), ~(\rm{\ref{P-RIS-1}}),~(\rm{\ref{P5-1-W_k}}),~(\rm{\ref{Rank-W}}), 
		\end{eqnarray}
	\end{subequations}
	
	The above problem (\ref{P5-4}) is an SDP problem, which can be efficiently solved by CVX.
	\begin{algorithm}[t]  
		\caption{SROCR-Based Algorithm for Solving (\ref{P5-4})}
		\label{Algorithm1}
		\renewcommand{\algorithmicrequire}{\textbf{Initialize}}
		\renewcommand{\algorithmicensure}{\textbf{Output}}
		\begin{algorithmic}[1]
			\STATE \textbf{Initialize} set iteration index $\tau=0$, the feasible point $\{\mathbf{W}_{c,k}^{(0)},w_{c,k}^{(0)}\}$, and the step size $\zeta^{(0)}$.
			\REPEAT 
			\STATE Obtain $\mathbf{W}_{c,k}$ by solving (\ref{P5-4});
			\IF {(\ref{P5-4}) is solvable} 
			\STATE Update $\mathbf{W}_{c,k}^{(\tau+1)} = \mathbf{W}_{c,k}$ and $\zeta^{(\tau+1)} =\zeta^{(0)}$;
			\ELSE {}
			\STATE Update $\mathbf{W}_{c,k}^{(\tau+1)} = \mathbf{W}_{c,k}^{(\tau)}$ and $\zeta^{(\tau+1)} =\zeta^{(\tau)} /2$;
			\ENDIF
			\STATE Update $\tau=\tau +1$;
			\STATE Update $w_{c,k}^{(\tau)} = \rm{min}(1, \frac{\lambda_{\rm{max}}(\mathbf{W}_{c,k}^{(\tau)})}{\mathrm{Tr}(\mathbf{W}_{c,k}^{(\tau)})}+\zeta^{(\tau)})$;
			\UNTIL the stopping criterion is satisfied.
		\end{algorithmic}
	\end{algorithm}	
	Algorithm \ref{Algorithm1} presents the procedure for solving problem (\ref{P5-3}), where the SROCR technique is used to gradually relax the rank-one constraint.
	
	\subsection{MF-RIS Coefficients Design}
	With fixed $\mathbf{W}_{c,k}$, we optimize $\mathbf{V}_{c}$. The MF-RIS coefficients subproblem is given by
	\begin{subequations}
		\label{P5-2}
		\begin{eqnarray}
			& \!\!\!\!\!\!\!\! \max \limits_{\mathbf{V}_c, \Omega} 
			\label{P5-2-function}
			&\!\!\!  \psi  \\
			\label{P5-2-1}
			&\!\!\!\!\!\!\!\!  \mathrm{s.t.} &\!\!\!\!  \widehat{E}_{c,k}  +2 \widehat{e}_{c,k}\le \widehat{x}_{c,k}, ~\forall c,~\forall k,\\
			\label{e_c}
			&&\!\!\!\! \left(\varpi_{h,c,k}^{2}\!+\!\varpi_{F,c,k}^{2}\!\! \sum\nolimits_{m}\! \beta_m^c\right)\mathrm{Tr}\left\{\mathbf{C}_{c,k}\right\} \!-\!\sigma_1^2\varpi_{f,c,k}^2\mathrm{Tr}\{\boldsymbol{\Phi}_{c}\} \nonumber \\
			&&\!\!\!\! - \sqrt{2\ln(1 / \rho_{c,k})}\widetilde{x}_{c,k} +\ln(\rho_{c,k})\widetilde{y}_{c,k} +{e}_{c,k} \ge 0, \\
			\label{P5-2-2}
			&&\!\!\!\! \widetilde{y}_{k}\mathbf{I}_{M+N}+\mathbf{B}_{c,k} \succeq \mathbf{0}, ~\widetilde{y}_{k} \ge 0, ~\forall c,~\forall m,~\forall k,  \\
			&&\!\!\!\! \mathrm{Rank}(\mathbf{V}_c) = 1, \forall c,~\forall k,  \\
			\label{P5-2-V_c}
			&&\!\!\!\! \mathbf{V}_c \succeq \mathbf{0}, [\mathbf{V}_c]_{m,m} = \beta_m^c,[\mathbf{V}_c]_{M+1,M+1} \!=\! 1,\forall c, \forall k,\\
			&&\!\!\!\! (\rm{\ref{P0-C-RIS-coefficients}}),~(\rm{\ref{P0-C-RIS}}),~ (\rm{\ref{P2-1}}),  ~(\rm{\ref{P2-r}}), ~(\rm{\ref{P-RIS-2}}), ~(\rm{\ref{P-RIS-1}}).   
		\end{eqnarray}
	\end{subequations}
	
	Similar to $\mathrm{Rank}({\mathbf{W}_{k}}) = 1$, for $\mathrm{Rank}(\mathbf{V}_c) = 1$,
	we have 
	\begin{align}
		\label{Rank-V}
		(\mathbf{v}_{c,k}^{(\tau-1)})^{\mathrm{H}}\mathbf{V}_{c,k}^{(\tau)}\mathbf{v}_{c,k}^{(\tau-1)} \ge v_{c,k}^{(\tau-1)} \mathrm{Tr}(\mathbf{V}_{c,k}^{(\tau)}),~ \forall c,~\forall k,
	\end{align}
	where $\mathbf{v}_{c,k}^{(\tau-1)}$ is the eigenvector corresponding to the largest eigenvalue of $\mathbf{V}_{c,k}^{(\tau-1)}$. Here $v_{c,k}^{(\tau-1)} \in (0,1]$ denotes the trace ratio parameter of $\mathbf{V}_{c,k}^{(\tau-1)}$, which is the solution obtained in the $(\tau-1)$-th iteration.
	Finally, problem (\ref{P5-2}) is given by
	\begin{subequations}
		\label{P5-3}
		\begin{eqnarray}
			& \!\!\!\!\!\!\!\!\!\!\!\!\! \max \limits_{\mathbf{V}_c, \Omega} 
			\label{P5-3-function}
			&\!\!\!\!  \psi  \\
			\label{P5-3-1}
			&\!\!\!\!\!\!\!\!\!\!\!\!\!  \mathrm{s.t.} &\!\!\!\!  \widetilde{E}_{c,k}  +2 \widetilde{e}_{c,k}\le \widehat{x}_{c,k},~\forall c,~\forall k, \\
			\label{P5-3-2}
			&&\!\!\!\! \left(\varpi_{h,c,k}^{2}\!+\!\varpi_{F,c,k}^{2}\! \mathrm{Tr}\{\boldsymbol{\Phi}_{c}\}\right)\mathrm{Tr}\left\{\mathbf{C}_{c,k}\right\} \!-\!\sigma_1^2\varpi_{f,c,k}^2\mathrm{Tr}\{\boldsymbol{\Phi}_{c}\} \nonumber \\
			&&\!\!\!\! - \sqrt{2\ln(1 / \rho_{c,k})}\widetilde{x}_{c,k} +\ln(\rho_{c,k})\widetilde{y}_{c,k} +\bar{e}_{c,k} \ge 0, \\
			\label{P5-3-3}
			&&\!\!\!\! \widetilde{y}_{c,k}\mathbf{I}_{M+N}+ \widetilde{\mathbf{B}}_{c,k} \succeq \mathbf{0},~\widetilde{y}_{k} \ge 0,~\forall c,~\forall k,\\
			&&\!\!\!\! (\rm{\ref{P0-C-RIS-coefficients}}),\!~(\rm{\ref{P0-C-RIS}}),\!~ (\rm{\ref{P2-1}}),  \!~(\rm{\ref{P2-r}}), \!~(\rm{\ref{P-RIS-2}}), \!~(\rm{\ref{P-RIS-1}}), \!~(\rm{\ref{P5-2-V_c}}),\!~(\rm{\ref{Rank-V}}).   
		\end{eqnarray}
	\end{subequations}
	where $\widetilde{E}_{c,k}$ and $\widetilde{e}_{c,k}$ are given in Appendix \ref{Appendix D},
	Here $\widetilde{\mathbf{B}}_{c,k} = 
	\begin{bmatrix}
		(\varpi_{h,c,k}^{2}+\varpi_{F,c,k}^{2}\mathrm{Tr}\{\boldsymbol{\Phi}_{c}\})\mathbf{C}_{c,k} & \mathbf{0}_{N \times M} \\
		\mathbf{0}_{M \times N} & -\sigma_1^2\varpi_{f,c,k}^2 \boldsymbol{\Phi}_{c} 
	\end{bmatrix}.$
	Problem (\ref{P5-3}) is an SDP problem and can be efficiently solved by CVX. Finally, the formulated problem (\ref{P4}) is addressed by solving subproblems (\ref{P5-4}) and (\ref{P5-3}) iteratively.
	
	\section{Convergence and Complexity Analysis} \label{Convergence}
	
	In this work, we introduce the bounded and the statistical CSI error models, where subproblems (\ref{P3-1}) and (\ref{P3-4}) are solved alternately for the bounded CSI error model, while subproblems (\ref{P5-4}) and (\ref{P5-3}) are solved alternately for the statistical CSI error model. 
	Each subproblem (\ref{P3-1}) and (\ref{P3-4}) under the bounded CSI error model, and each subproblem (\ref{P5-4}) and (\ref{P5-3}) under the statistical CSI error model converges to their individual suboptimal solutions. 
	
	For the convergence analysis of the bounded and the statistical CSI error models, we have:
	\begin{align}
		\!\!\!\!\!\!\mathrm{EE}_{b}\Big(\mathbf{w}_{c,k}^{(\tau)}, \boldsymbol{\Theta}_c^{(\tau)}\Big) &\overset{(a)}{\operatorname*{\leq}} \mathrm{EE}_{b}\Big(\mathbf{w}_{c,k}^{(\tau+1)},\boldsymbol{\Theta}_c^{(\tau)}\Big) \nonumber \\
		&\overset{(b)}{\operatorname*{\leq}}
		\mathrm{EE}_{b}\Big(\mathbf{w}_{c,k}^{(\tau+1)},\boldsymbol{\Theta}_c^{(\tau+1)}\Big), \\
		\!\!\!\!\!\!\mathrm{EE}_{s}\Big(\mathbf{W}_{c,k}^{(\tau)}, \boldsymbol{\Theta}_c^{(\tau)}\Big) &\overset{(c)}{\operatorname*{\leq}} \mathrm{EE}_{s}\Big(\mathbf{W}_{c,k}^{(\tau+1)},\boldsymbol{\Theta}_c^{(\tau)}\Big) \nonumber \\
		&\overset{(d)}{\operatorname*{\leq}}
		\mathrm{EE}_{s}\Big(\mathbf{W}_{c,k}^{(\tau+1)},\boldsymbol{\Theta}_c^{(\tau+1)}\Big),
	\end{align}
	where $\tau$ denotes the iteration index. $\mathrm{EE}_b$ and $\mathrm{EE}_s$ represent the EE under the bounded and the statistical CSI error models, respectively. Here, $(a)$ and $(b)$ follow since the updates of $\mathbf{w}_{c,k}$ and  $\boldsymbol{\Theta}_c$ maximize EE when the other variables are fixed in the bounded CSI error model.
	Then, $(c)$ and $(d)$ are due to the updates of $\mathbf{W}_{c,k}$ and  $\boldsymbol{\Theta}_c$ to maximize EE when the other variables are fixed in the statistical CSI error model.
	Moreover, considering that $\mathrm{EE}_b(\mathbf{w}_{c,k},\boldsymbol{\Theta}_c)$ and $\mathrm{EE}_s(\mathbf{W}_{c,k},\boldsymbol{\Theta}_c)$ are monotonically non-decreasing over each iteration and its value is upper bounded, both algorithms converge to suboptimal solutions of the original problems (\ref{P0}) and (\ref{P4}) after several iterations.
	\begin{table}[t]
		\centering
		\caption{LMI and SOC constraints description}
		\scalebox{0.9}{\begin{tabular}{|c|c|c|c|c|c|} 
				\hline 
				Constraint   & Size & Number & Constraint  & Size & Number \\ 
				\hline
				(\rm{\ref{SINR1-2}})   & $a_1$ & $K$ & (\rm{\ref{amplification power2}}),~(\rm{\ref{P-RIS1}})  & $a_2$ & $1$ \\
				\hline
				(\rm{\ref{SINR2-1}})  & $a_3$ & $K$ & (\rm{\ref{SINR3-1}})   & $a_4$ & $K$ \\
				\hline
				(\rm{\ref{P5-4-E}}),~(\rm{\ref{P5-4-1}})  & $a_5$ & $K$ & (\rm{\ref{P5-4-2}})   & $v_1$ & $K$ \\
				\hline
				(\rm{\ref{P5-3-1}}),~(\rm{\ref{P5-3-2}})   & $a_6$ & $K$ & (\rm{\ref{P5-3-3}})  & $v_1$ & $K$ \\		
				\hline
		\end{tabular}}
		\label{LMI}
		\vspace{-3mm}
	\end{table}
	Since all the subproblems involving LMIs, second-order cones (SOCs), and linear constraints that can be efficiently tackled using a standard interior point method \cite{Boyd2004CVX}, we estimate the overall complexity based on their worst-case runtime. After ignoring the non-dominated linear constraints, the general expressions of computational complexity of different methods are expressed as \cite{Zhou2020CSI}
	\begin{align}
		&\!\!\!\!\! \mathcal{O}((\sum_{j=1}^Ja_j \!+\! 2I)^{1/2}n\underbrace{(n^2\!+\! n\sum_{j=1}^Ja_j^2 \!+ \! \sum_{j=1}^Ja^3}_{\text{due to LMIs}} \! +\! \underbrace{n\sum_{i=1}^Iv_i^2}_{\text{due to SOCs}})),
	\end{align}
	where $n$, $J$, and $I$ denote the number of variables, the number of LMIs with size $a_j$, and the number of SOC constraints with size $v_i$, respectively.
	
	For the bounded CSI error model, we optimize subproblems (\ref{P3-1}) and (\ref{P3-4}) alternately to solve the original problem (\ref{P0}) until convergence.
	The number and size of LMI and SOC constraints associated in subproblems (\ref{P3-1}) and (\ref{P3-4}) are presented in Table \ref{LMI}, where $a_1=MN+N+1$, $a_2=2MN+1$, $a_3=2N+K$, and $a_4=2M+1$.
	Therefore, the complexity of solving problem (\ref{P3-1}) and (\ref{P3-4}) are denoted by $\mathcal{O}_{\mathbf{w}}$ and $\mathcal{O}_{\mathbf{\Theta}}$, where 
	\begin{subequations}
		\begin{align}
			\!\!\!\! &\mathcal{O}_{\mathbf{w}} = \mathcal{O}(\sqrt{f_1}n_1(n_1^2+n_1f_2+f_3)), \\
			\!\!\!\! &\mathcal{O}_{\mathbf{\Theta}} = \mathcal{O}(\sqrt{f_1+4M}n_2(n_2^2+n_2f_2 +f_3+2n_2M)),\\
			\!\!\!\! &f_1 = K(a_1+a_3+a_4)+2a_2, ~n_1 = NK, ~n_2 = 2M, \\
			\!\!\!\! &f_2\!=\! K(a_1^2 \!+\! a_3^2\! +\! a_4^2) \!+\! 2a_2^2,\!~f_3 \!=\! K(a_1^3 \!+\! a_3^3\!+\! a_4^3) \!+ \!2a_1^3.
		\end{align}
	\end{subequations}
	
	For the statistical CSI error model, subproblems (\ref{P5-4}) and (\ref{P5-3}) are optimized alternately to solve the original problem (\ref{P4}) until the convergence threshold is satisfied.
	Table \ref{LMI} provides the number and size of LMI and SOC constraints contained in subproblems (\ref{P5-4}) and (\ref{P5-3}), where $a_5=N$, $a_6=M$, $v_1=(M+N)(M+N+1)$.
	Therefore, the complexity of solving problem (\ref{P5-4}) and (\ref{P5-3}) are denoted by $\mathcal{O}_{\mathbf{W}}$ and $\mathcal{O}_{\mathbf{V}}$, where
	\begin{subequations} 
		\begin{align}
			\!\!\!\!\!\!\!\!\!\! \mathcal{O}_{\mathbf{W}}    =& \mathcal{O}(\sqrt{2K(a_5+1)}n_3(n_3^2+2n_3Ka_5^2+2Ka_5^3 \nonumber \\
			& +n_3Kv_1^2)),~n_3 = NK, \\
			\!\!\!\!\!\!\!\!\!\! \mathcal{O}_{\mathbf{V}} = &\mathcal{O}(\sqrt{2K(a_6+1)}n_4(n_4^2+2n_4Ka_6^2+2Ka_6^3 \nonumber \\
			& +n_4Kv_1^2),~n_4 = 2M.
		\end{align}
	\end{subequations}
	
	\vspace{-2mm}
	\section{Simulation Results} \label{Simulation} 
	\vspace{-1mm}
	In this section, we provide simulation results to evaluate the performance of the proposed algorithms. 
	The BS and the MF-RIS are located at $(0,0,10)$ and $(0,35,20)$ meter (m), respectively. The users are uniformly distributed on circles of radius $3$ m, and centered at $(0,30,0)$ and $(0,40,0)$ m for the reflection and refraction spaces, respectively.  
	The path loss of all channels is presented as $\mathrm{PL} = -\mathrm{PL_0} -10 \alpha\mathrm{log}_{10}(d)$ dB, where $d$ is the link distance in m and $\alpha$ denotes the path loss exponent. $\mathrm{PL_0} = 30$ dB is the signal attenuation at the distance of 1 m \cite{Ma2024}.
	In addition, we assume that all channels follow Rician fading distribution with Rician factor of $3$ dB, which are given by 
	\begin{subequations}
		\begin{align}
			&\mathbf{{G}} = \sqrt{PL(d_G)}(\sqrt{\frac{\kappa}{1+\kappa}}\mathbf{G}^{\mathrm{LOS}}+\sqrt{\frac{1}{1+\kappa}}\mathbf{G}^{\mathrm{NLOS}}),\\
			&\mathbf{{h}}_{c,k} = \sqrt{PL(d_h)}(\sqrt{\frac{\kappa}{1+\kappa}}\mathbf{h}_{c,k}^{\mathrm{LOS}}+\sqrt{\frac{1}{1+\kappa}}\mathbf{h}_{c,k}^{\mathrm{NLOS}}),\\
			&\mathbf{{f}}_{c,k} = \sqrt{PL(d_f)}(\sqrt{\frac{\kappa}{1+\kappa}}\mathbf{f}_{c,k}^{\mathrm{LOS}}+\sqrt{\frac{1}{1+\kappa}}\mathbf{f}_{c,k}^{\mathrm{NLOS}}),
		\end{align}
	\end{subequations}
	where $\kappa$ is the Rician factor. Here $\mathbf{G}^{\mathrm{LOS}}$, $\mathbf{h}_{c,k}^{\mathrm{LOS}}$, and $\mathbf{f}_{c,k}^{\mathrm{LOS}}$ stand for the deterministic line-of-sight (LOS) component. $\mathbf{G}^{\mathrm{NLOS}}$, $\mathbf{h}_{c,k}^{\mathrm{NLOS}}$, and $\mathbf{f}_{c,k}^{\mathrm{NLOS}}$ stand for the deterministic non-LOS (NLOS) component, which has i.i.d circularly symmetric complex Gaussian (CSCG) entries with zero mean and unit variance. $PL(d_G)$, $PL(d_h)$, and $PL(d_f)$ are the path loss for BS-RIS, MF-RIS-user, and BS-user links, respectively.
	Furthermore, we set the total power budget as $P_{T}^{\mathrm{max}}$, where $P_{\mathrm{RIS}}^{\mathrm{max}} = P_{\mathrm{BS}}^{\mathrm{max}} = P_{T}^{\mathrm{max}}/2$ for the MF-RIS-assisted multi-user system \cite{Jiang2023-RIS,Ren2023}, and $P_{\mathrm{BS}}^{\mathrm{max}} = P_{T}^{\mathrm{max}}$ for the passive RIS (No RIS)-aided networks.

	For comparison, we provide the following benchmarks: 
	\begin{itemize}
		\item
		\textbf{STAR-RIS}: The STAR-RIS reflects and refracts the incoming signals simultaneously without amplification, i.e., $\beta_{\rm max} = 1$.
		\item
		\textbf{Active RIS}: The active RIS simultaneously supports signal reflection and amplification. i.e., $\boldsymbol{\Theta}_t = \mathbf{0}_{M \times M}$.
		\item 
		\textbf{SF-RIS}: In this scheme, the SF-RIS only supports signal reflection, i.e.,  $\beta_{\rm max} = 1$, and $\boldsymbol{\Theta}_t = \mathbf{0}_{M \times M}$.
		\item
		\textbf{No-RIS}: The RIS is not deployed in this scheme. Only the channels between BS and users are available.
\end{itemize}

	Based on \cite{Wang2014CSI}, 
	we define $\varpi_{h,c,k}^2=\delta_{h,c,k}^2 \Vert \widehat{\mathbf{h}}_{c,k} \Vert_2^2$, $\varpi_{f,c,k}^2=\delta_{f,c,k}^2 \Vert \widehat{\mathbf{f}}_{c,k} \Vert_2^2$, and $\varpi_{G}^2=\delta_{G}^2 \Vert \mathrm{vec}(\widehat{\mathbf{G}}) \Vert_2^2$ for the statistical CSI error model, respectively, where $\delta_{h,c,k} \in [0,1)$, $\delta_{f,c,k} \in [0,1)$, and $\delta_{G} \in [0,1)$ are used to scale the level of CSI uncertainties. For
	$\mathbf{F}_{c,k}=\mathrm{diag}(\mathbf{f}_{c,k}^{\mathrm{H}})\mathbf{G}$, we obtain $\varpi_{F,c,k} = \varpi_{G} \Vert \mathrm{diag}(\widehat{\mathbf{f}}_{c,k}^{\mathrm{H}}) \Vert_F + \varpi_{f,c,k} \Vert \widehat{\mathbf{G}} \Vert_F +\varpi_{G}\varpi_{f,c,k} = \frac{(\varpi_{G}\varpi_{f,c,k})(\delta_{f,c,k}+\delta_{G}+\delta_{G}\delta_{f,c,k})}{\delta_{G}\delta_{f,c,k}} $.
	
	For the bounded CSI error model, we model the radii of the uncertainty regions as
	\begin{subequations}
		\begin{align}
			& \xi_{h,c,k} \!=\! \sqrt{\frac{\varpi_{h,c,k}^2}{2}{F}_{2N}^{-1}(1-\rho_q)},\\
			&\xi_{f,c,k} \!=\! \sqrt{\frac{\varpi_{f,c,k}^2}{2}{F}_{2M}^{-1}(1-\rho_q)},\\
			& \xi_{F,c,k} \!=\! \sqrt{\frac{\varpi_{F,c,k}^2}{2}{F}_{2MN}^{-1}(1-\rho_q)},\\
			&\xi_{G} \!=\! \sqrt{\frac{\varpi_{G}^2}{2}{F}_{2MN}^{-1}(1-\rho_q)},
		\end{align}
	\end{subequations}
	where ${F}_{2N}^{-1}(\centerdot)$, ${F}_{2M}^{-1}(\centerdot)$, and ${F}_{2MN}^{-1}(\centerdot)$ denote the inverse cumulative distribution functions of the Chi-square distribution with degrees of freedom equal to $2N$, $2M$, and $2MN$, respectively. 
	\begin{table*}[t]
		\centering
		\caption{Simulation Parameters}
		\scalebox{0.87}{\begin{tabular}{|c|c|} 
				\hline
				Parameter   & Value  \\ 
				\hline
				Pathloss exponents of BS-RIS, RIS-user, and BS-user links   & 2.2, 2.6, 2.8   \\
				\hline
				Noise power at the MF-RIS and users  & $\sigma_1^2=\sigma_2^2=-80~\mathrm{dBm}$  \\
				\hline
				\multirow{2}{*}{Power consumption parameters}   & 	$P_{T}^{\mathrm{max}}=30~\mathrm{dBm}$,  $P_{\mathrm{PS}}=-10~\mathrm{dBm}$, 
				$P_{\mathrm{PA}}=-5~\mathrm{dBm}$ \cite{Long2021-activeRIS,Zhi2022},\\
				& 
				$P_{\mathrm{U}}=10~\mathrm{dBm}$ \cite{Wang2022CSI}, $P_{\rm S}=40~\mathrm{dBm}$,  $P_{\mathrm{RF}}=30~\mathrm{dBm}$ \cite{Niu2024}, $\xi = \zeta = 1.1$.  \\	
				\hline
				Initial penalty factor   & $\lambda^{(0)}= 10^{-3}$     \\
				\hline
				Maximum outage probability   & $\rho=0.05$    \\
				\hline
				Minimum rate requirement   & $R_{c,k}^{\mathrm{min}} = 1~\mathrm{bits/s/Hz}$    \\
				\hline
				Scaling factor   & $\varepsilon$ = 10    \\
				\hline
				Convergence tolerance   & $10^{-4}$    \\	
				\hline
				Other parameters   & $N=6, M=32, K_r = K_t = 3$    \\	
				\hline
		\end{tabular}}
		\label{Para}
	\end{table*}
	We set $\delta_{h}^2 = 0.02$ and $\delta_{G}^2 = \delta_{f}^2 = 0.01$, $\forall c,~k$, which represents a case where the CSI uncertainty of the direct link is higher than that of the cascade link \cite{Zhou2020CSI}.
	Note that since the MF-RIS amplifies noise while amplifying desired signals, the CSI uncertainty of the cascade link may be higher than that of the direct link. Other simulation settings are listed in Table \ref{Para}.
	
	\begin{figure}[t]
		\centering
		\includegraphics[width=3 in]{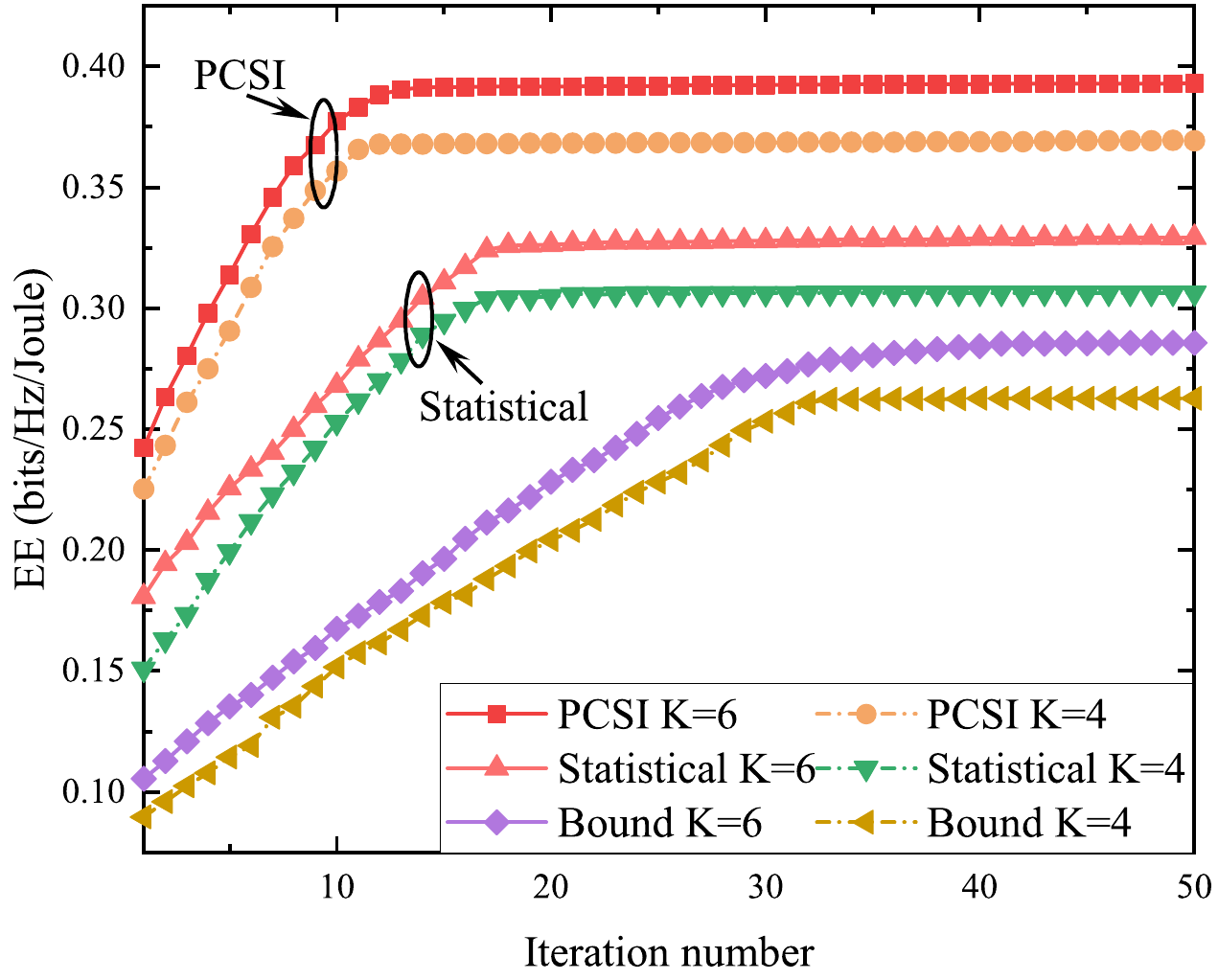}
		\caption{Convergence performance of the proposed algorithm. For the perfect CSI, i.e., PCSI, scenario, we set $\delta_{h} = \delta_{G} = \delta_{f} = 0$ in the statistical CSI error model.}
		\label{Iteration}
		\vspace{-5mm}
	\end{figure}
	
	The convergence behaviors of the proposed algorithms when considering different CSI error models are depicted in Fig. \ref{Iteration}. We observe that all curves exhibit a gradual increase and eventually converge after a certain number of iterations. Moreover, the proposed algorithms under the perfect CSI scenario converge faster than those with imperfect CSI scenarios since the CSI uncertainty increases the dimension of the optimization problem. Specifically, more additional iterations are required for convergence with the increase of $K$. It is reasonable that as $K$ increases, both the dimensions of LMI constraints and the number of optimization variables also increase. Furthermore, the algorithms under the bounded CSI error model need more iterations to converge compared to those under the statistical CSI error model.
	
	\begin{figure}[t]
		\centering
		\includegraphics[width=3 in]{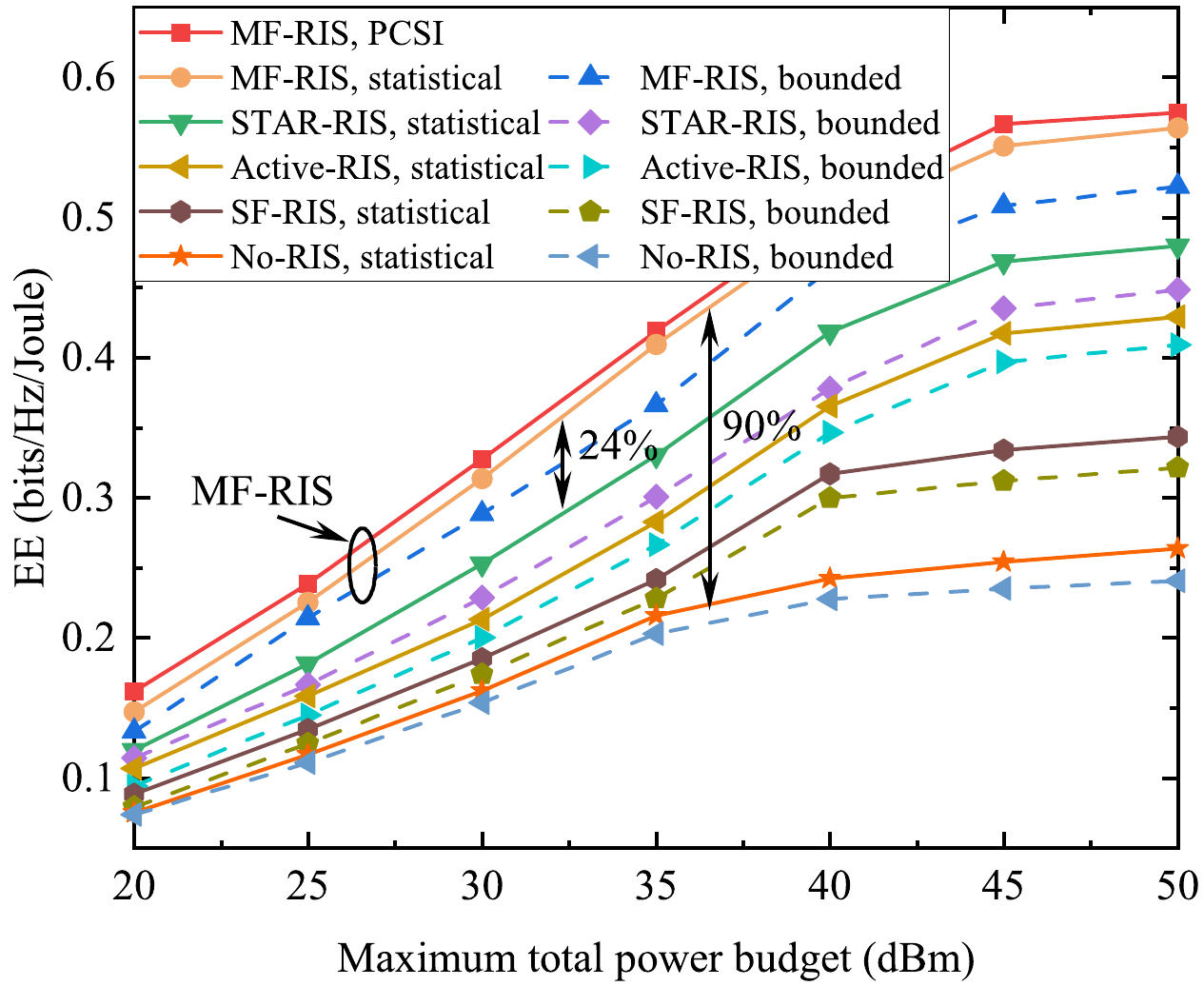}
		\caption{EE versus the transmit power budget.}
		\label{P}
		\vspace{-5mm}
	\end{figure}
	
	Fig. \ref{P} investigates the system EE versus the maximum transmit power budget $P_{T}^{\mathrm{max}}$. From this figure, it is obvious that EE exhibits an increasing trend with $P_{T}^{\mathrm{max}}$ until a certain threshold is reached. Then, EE keeps a stable level as $P_{T}^{\mathrm{max}}$ exceeds the threshold value for all schemes. This can be explained as follows. When $P_{T}^{\mathrm{max}}$ is low, full transmit power is exploited for data transmission. Once the maximum EE is achieved with a large $P_{T}^{\mathrm{max}}$, although $P_{T}^{\mathrm{max}}$ still rises, the actual transmit power remains constant. In this case, EE stabilizes.
	In addition, it is seen that the EE of the statistical CSI error model is larger compared to that of the bounded CSI error model. The reason is that the bounded CSI error model is a more conservative robust design. Therefore, more transmit power are needed to guarantee that each user reaches the desired rate requirement, resulting in a lower EE. Furthermore, as shown in Fig. \ref{P}, the MF-RIS scheme enjoys the best performance compared to all other schemes. Specifically, for the statistical CSI error model with $P_{T}^{\mathrm{max}} = 35~\mathrm{dBm}$, the MF-RIS scheme can enjoy up to 24\% and 90\% higher EE compared to the STAR-RIS and No-RIS schemes, respectively. The results reveal that the MF-RIS can effectively address the half-space coverage and overcome the double-fading attenuation problem faced by the SF-RIS and the STAR-RIS. Moreover, the STAR-RIS scheme is superior to the active RIS scheme. The reason is that the STAR-RIS can serve both reflective and refractive users to improve the system performance. In contrast, the active RIS requires extra energy consumption to enable the operation of the amplifier circuit, and can only achieve signal reflection.
	
	\begin{figure}[t]
		\centering
		\includegraphics[width=3 in]{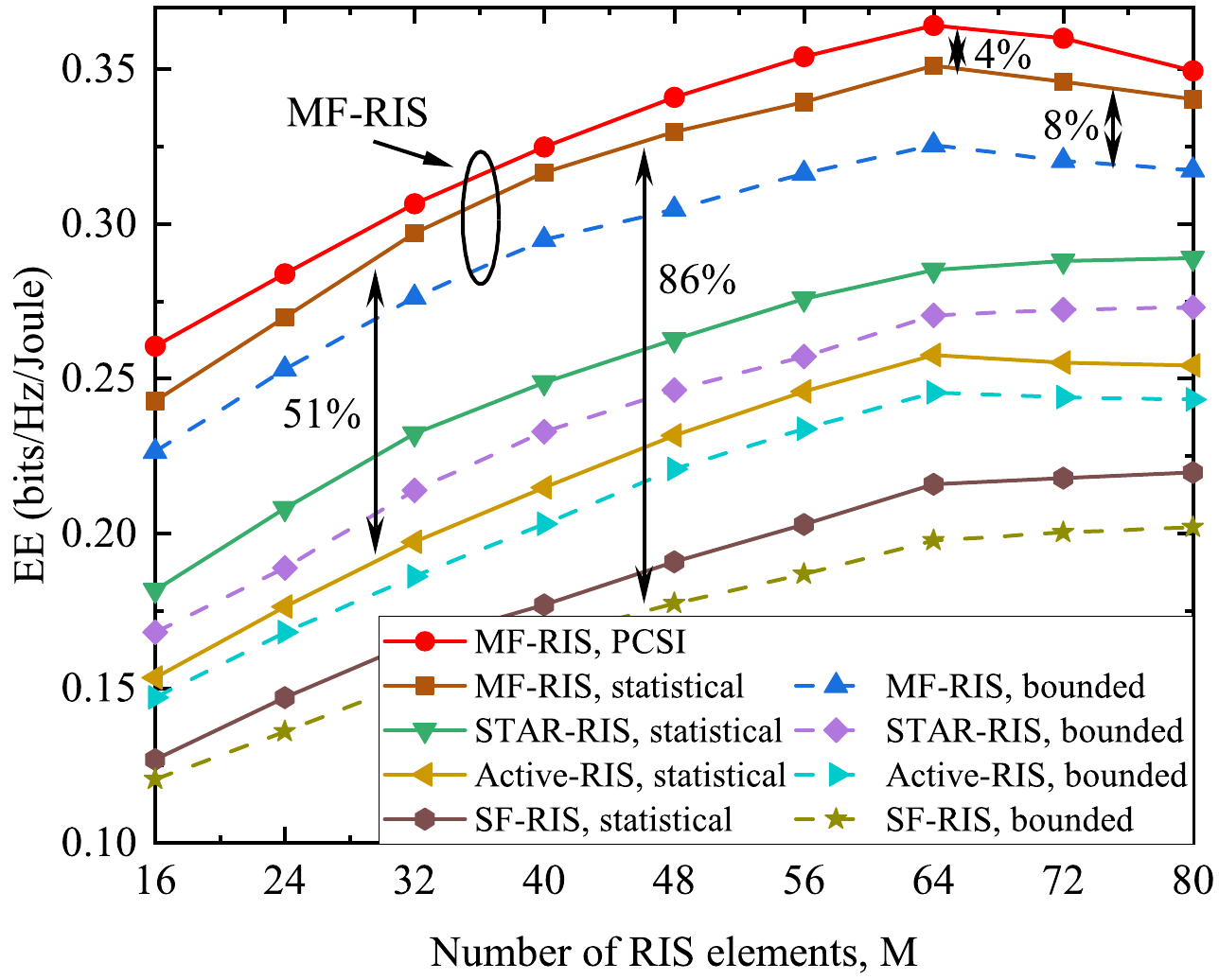}
		\caption{EE versus the number of RIS elements.}
		\label{M}
		\vspace{-5mm}
	\end{figure}
	
	Fig. \ref{M} shows the EE versus the number of RIS elements $M$. As observed in Fig. \ref{M}, the MF-RIS scheme with perfect CSI has 4\% EE performance gain compared with the MF-RIS scheme with statistical CSI error model. This is reasonable because the scheme with perfect CSI can take full advantage of the available channel conditions without any estimation error. Since the EE performance gap between the scheme with perfect CSI and the scheme with CSI errors is small, the effectiveness of the proposed algorithm can be demonstrated.
	Moreover, the MF-RIS scheme obtains the best performance. Specifically, the MF-RIS is able to enjoy a 86\% higher EE compared to SF-RIS scheme under the statistical CSI error model when $M=48$. Moreover, the EE of the MF-RIS scheme under the statistical CSI error model outperforms that under the bounded CSI error model with a 8\% higher EE. In addition,
	it is obvious that the EE of MF-RIS and active RIS schemes first increases with $M$, and then decreases after reaching a peak value. This is because when $M$ is small, the spectral efficiency increases significantly as $M$ increases, thus leading to an improvement of EE. However, with the further increase of $M$, the energy consumed in controlling phase shift circuit and the DC biasing of the amplifier circuit for all elements increases, leading to a tremendous increase in the total energy consumption of MF-RIS. In this case, the total power consumed grows faster than spectral efficiency, thereby resulting in EE degradation.
	The reason for the similar trend of active RIS scheme is the same as that of MF-RIS scheme. Therefore, it is crucial to design the number of elements in actual deployment to meet green communication requirements.
	Furthermore, although the EE performance of STAR-RIS and SF-RIS schemes continues to improve over a wide range of $M$, the growth rate gradually slows as the static power consumption, e.g., the phase shift circuit, by each element accumulates.
	
	\begin{figure}[t]
		\centering
		\includegraphics[width=3 in]{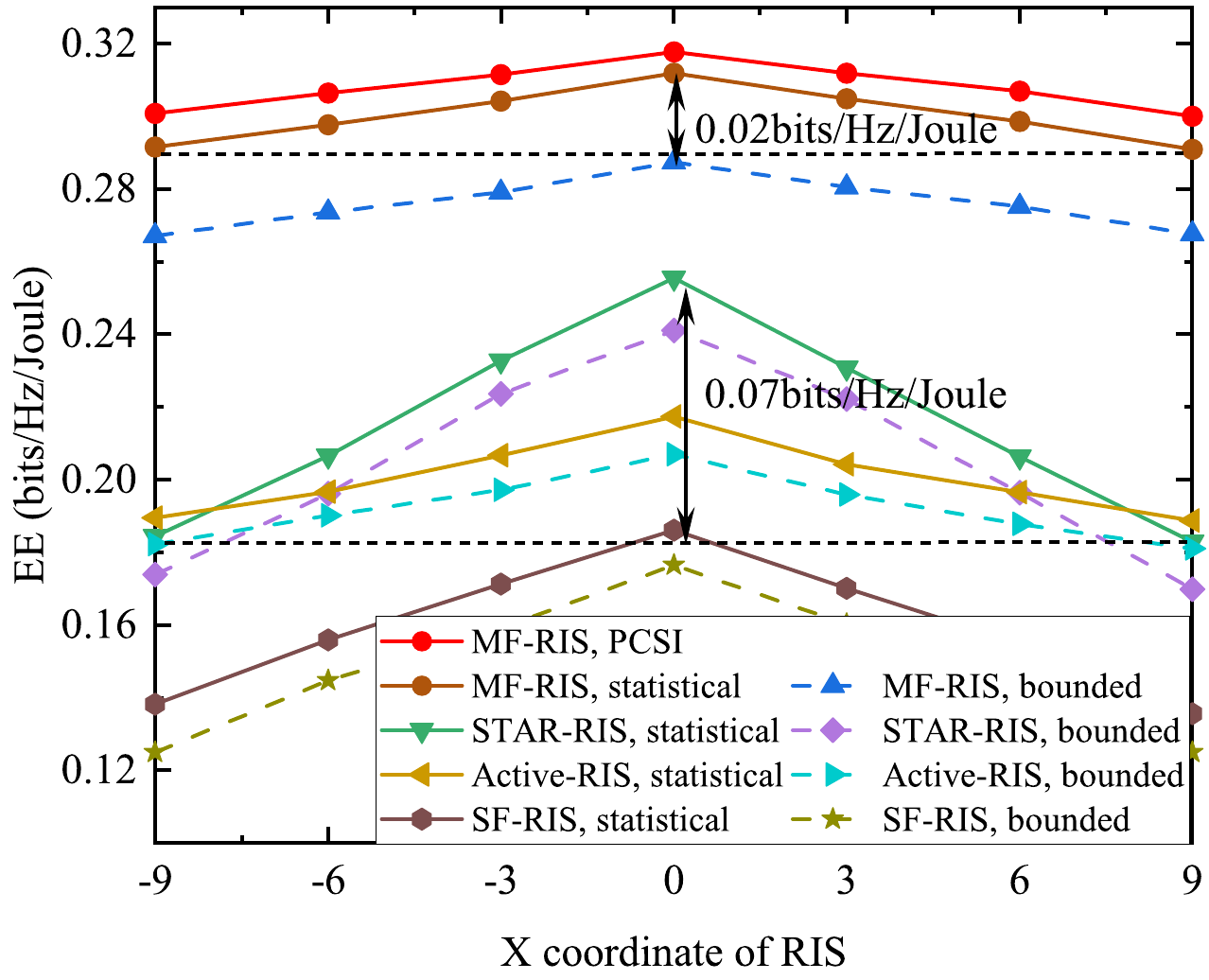}
		\caption{EE versus the X-coordinate of RIS.}
		\label{X}
		\vspace{-5mm}
	\end{figure}
	
	Fig. \ref{X} presents the effect of RIS location on the EE performance. In this figure, it is apparent that the EE of MF-RIS consistently outperforms all other counterparts as the RIS moves. Moreover, all schemes obtain the best performance when $x \! =\!0$. Furthermore, as the distance from $x \!=\!0$ increases, the EE performance of all schemes decreases. This is because the channel gain decreases with the increase of link distance. Particularly, as the distance of BS-MF-RIS and MF-RIS-user increase, the signal attenuation is intensified and the system spectral efficiency decreases, resulting in a decrease in EE. In addition, compared to STAR-RIS and SF-RIS schemes, the MF-RIS and active RIS schemes have a more stable performance as the RIS location changes. Specifically, when the RIS location changes from $x=0$ to $x=-9$, the EE performance of MF-RIS and STAR-RIS schemes decrease by 0.02 bits/Hz/Joule and 0.07 bits/Hz/Joule, respectively. This reveals the amplification function of MF-RIS can alleviate the double-fading attenuation and provide a stable system performance.
	
	\begin{figure}[t]
		\centering
		\includegraphics[width=3 in]{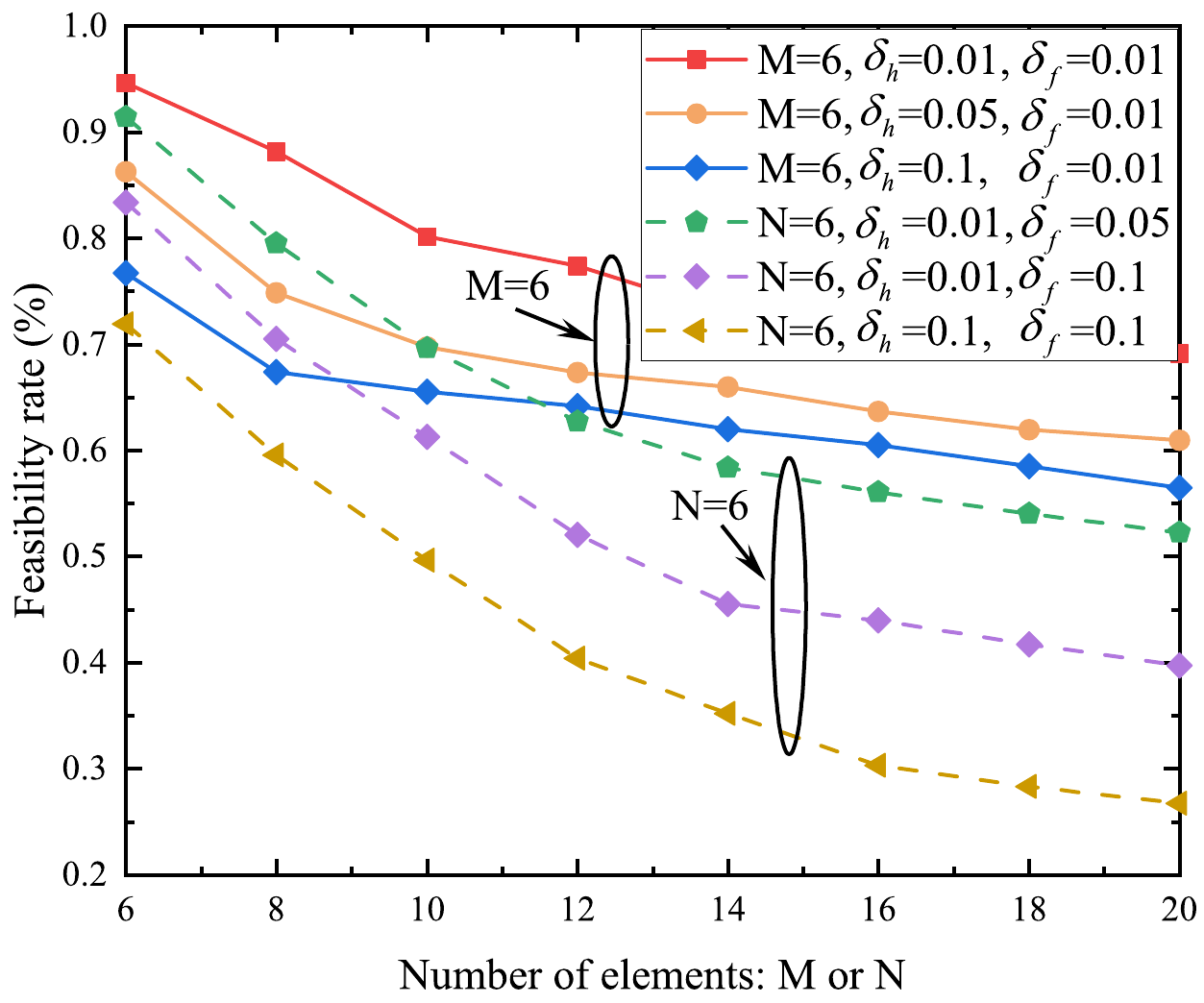}
		\caption{Feasibility rate versus the number of elements.}
		\label{Fea}
		\vspace{-5mm}
	\end{figure}
	
	Fig. \ref{Fea} depicts the feasibility rate versus the number of RIS elements $M$ and BS antennas $N$ for various levels of channel uncertainty $\delta_{h}$ and $\delta_{f}$, where $\delta_{G} = 0.01$. Specifically, the feasibility rate is defined as the ratio of the number of feasible channel realizations to the total number of channel realizations, where the feasible channel realization means that there exists a feasible solution to the outage constrainted problem in problem (\ref{P4}) with these channel realizations \cite{Zhou2020CSI}. As shown in Fig. \ref{Fea}, whether $N$ or $M$ is fixed, an increase in the channel uncertainty levels $\delta_{h}$ and $\delta_{f}$ leads to a decrease in the feasibility rate. It is intuitive as a larger $M$ or $N$ will introduce more channel estimation errors under high $\delta_{h}$ and $\delta_{f}$, thus resulting in a lower feasibility rate. Moreover, when $N$ is fixed, the decreasing trend of feasibility rate with increasing $M$ is greater than that of with increasing $N$ when $M$ is fixed. This is because more degree of freedoms are provided as $N$ increases, thereby compensating the channel estimation error.
	
	\vspace{-2mm}
	\section{Conclusion} \label{Conclusion}
	\vspace{-1mm}
	We explored robust beamforming designs under imperfect CSI for the MF-RIS-assisted multiuser wireless system in this paper. Specifically, we studied the EE maximization problem under the bounded and the statistical CSI error models. To address the CSI uncertainty, we exploited the S-procedure and the Bernstein-Type Inequality to tackle the QoS constraint in the bounded CSI error model, and the rate outage probability constraint in the statistical CSI error model, respectively. Then, for each model, we applied an AO framework to optimize the MF-RIS coefficients and the BS transmit beamforming vector alternately, where SCA, SDR, PCCP, and SROCR methods are utilized to solve each sub-problem. Numerical results verified the convergence and effectiveness of the proposed algorithms, and the significant potential of MF-RIS in energy-efficient communications, where MF-RIS enjoys 24\% , 38\%, and 79\% higher EE than STAR-RIS, active RIS, and SF-RIS schemes, respectively.
	In addition, our proposed algorithms revealed that the selection of MF-RIS size in the actual deployment is crucial to meet green communication requirements. Moreover, increasing the number of RIS elements results in a greater cumulative CSI error compared to increasing the number of BS antennas.
	
	\section*{Acknowledgement}
	The authors would like to express their sincere gratitude to the Editor and the anonymous Reviewers for their insightful comments and constructive suggestions, which have significantly improved the quality and clarity of this paper. Additionally, the work of Dr. Wanli Ni presented in this paper, especially the core concept of MF-RIS, was conducted during his doctoral studies at Beijing University of Posts and Telecommunications.
	
	\appendices
	\vspace{-2mm}
	\section{The Derivation of (\ref{eqn16})} \label{Appendix A}
	\vspace{-1mm}
	Constraint (\ref{eqn16}) is derived as follows:
	\begin{subequations}
		\begin{align}
			&\!\!\!\! \mathbf{A}_{c,k}=\widehat{\mathbf{A}}_{c,k}+\widehat{\mathbf{A}}_{c,k}^{\rm H}-\widetilde{\mathbf{A}}_{c,k}, \mathbf{a}_{c,k}=\widehat{\mathbf{a}}_{c,k}+\widetilde{\mathbf{a}}_{c,k}-\bar{\mathbf{a}}_{c,k}, \\
			&\!\!\!\! \widehat{\mathbf{A}}_{c,k} \!\!= \!\!
			\begin{bmatrix}
				\mathbf{w}_{c,k}^{(\tau)} \\
				\mathbf{w}_{c,k}^{(\tau)} \otimes \mathbf{u}_c^{(\tau),\ast}
			\end{bmatrix}
			\begin{bmatrix}
				\mathbf{w}_{c,k}^{\rm H}  \quad \mathbf{w}_{c,k}^{\rm H}  \otimes \mathbf{u}_c^{\rm T}
			\end{bmatrix},\\
			&\!\!\!\! \widetilde{\mathbf{A}}_{c,k} \!\!=  \!\!
			\begin{bmatrix}
				\mathbf{w}_{c,k}^{(\tau)} \\
				\mathbf{w}_{c,k}^{(\tau)} \otimes \mathbf{u}_c^{(\tau),\ast}
			\end{bmatrix}
			\begin{bmatrix}
				\mathbf{w}_{c,k}^{(\tau),{\rm H}}  \quad \mathbf{w}_{c,k}^{(\tau),{\rm H}}  \otimes \mathbf{u}_c^{(\tau),{\rm T}}
			\end{bmatrix}, \\
			&\!\!\!\! \widehat{\mathbf{a}}_{c,k}\!\!=\!\!
			\begin{bmatrix}
				\mathbf{w}_{c,k} \mathbf{w}_{c,k}^{(\tau),{\rm H}}(\widehat{\mathbf{h}}_{c,k}^{\rm H}+\widehat{\mathbf{F}}_{c,k}^{\rm H}\mathbf{u}_c^{(\tau)}) \\
				\mathrm{vec}^{\ast}(\mathbf{u}_c(\widehat{\mathbf{h}}_{c,k}+\mathbf{u}_c^{(\tau),{\rm H}}\widehat{\mathbf{F}}_{c,k})\mathbf{w}_{c,k}^{(\tau)}\mathbf{w}_{c,k}^{\rm H})
			\end{bmatrix}, \\
			&\!\!\!\! \widetilde{\mathbf{a}}_{c,k}\!\!=\!\!
			\begin{bmatrix}
				\mathbf{w}_{c,k}^{(\tau)} \mathbf{w}_{c,k}^{\rm H} (\widehat{\mathbf{h}}_{c,k}^{\rm H}+\widehat{\mathbf{F}}_{c,k}^{\rm H}\mathbf{u}_c) \\
				\mathrm{vec}^{\ast}(\mathbf{u}_c^{(\tau)}(\widehat{\mathbf{h}}_{c,k}+\mathbf{u}_c^{\rm H}\widehat{\mathbf{F}}_{c,k})\mathbf{w}_{c,k}\mathbf{w}_{c,k}^{(\tau),{\rm H}})
			\end{bmatrix}, \\
			&\!\!\!\!  \bar{\mathbf{a}}_{c,k}\!\!=\!\!
			\begin{bmatrix}
				\mathbf{w}_{c,k}^{(\tau)} \mathbf{w}_{c,k}^{(\tau),{\rm H}} (\widehat{\mathbf{h}}_{c,k}^{\rm H}+\widehat{\mathbf{F}}_{c,k}^{\rm H}\mathbf{u}_c^{(\tau)}) \\
				\mathrm{vec}^{\ast}(\mathbf{u}_c^{(\tau)}(\widehat{\mathbf{h}}_{c,k}+\mathbf{u}_c^{(\tau),{\rm H}}\widehat{\mathbf{F}}_{c,k})\mathbf{w}_{c,k}^{(\tau)}\mathbf{w}_{c,k}^{(\tau),{\rm H}})
			\end{bmatrix}, \\
			&\!\!\!\!  \widetilde{a}_{c,k} \!\!= \!\! (\widehat{\mathbf{h}}_{c,k}+\mathbf{u}_c^{(\tau),{\rm H}}\widehat{\mathbf{F}}_{c,k})\mathbf{w}_{c,k}^{(\tau)}\mathbf{w}_{c,k}^{\rm H}(\widehat{\mathbf{h}}_{c,k}^{\rm H}+\widehat{\mathbf{F}}_{c,k}^{\rm H}\mathbf{u}_c), \\
			&\!\!\!\! \widehat{a}_{c,k} \!=\!(\widehat{\mathbf{h}}_{c,k} \! +\! \mathbf{u}_c^{(\tau),{\rm H}}\widehat{\mathbf{F}}_{c,k})\mathbf{w}_{c,k}^{(\tau)}\mathbf{w}_{c,k}^{(\tau),{\rm H}}(\widehat{\mathbf{h}}_{c,k}^{\rm H} \!+\! \widehat{\mathbf{F}}_{c,k}^{\rm H}\mathbf{u}_c^{(\tau)}),\\
			&\!\!\!\! a_{c,k} \!=\!  2\Re{(\widetilde{a}_{c,k})} \!-\! \widehat{a}_{c,k},~\mathbf{x}_{c,k} \!=\! [\Delta\mathbf{h}_{c,k} \!\quad\! \mathrm{vec}^{\rm H}(\Delta\mathbf{F}_{c,k}^{\ast})]^{\rm H}. 
		\end{align}
	\end{subequations}
	
	\vspace{-2mm}
	\section{The Derivations of (\ref{SINR2}) and (\ref{SINR3})} \label{Appendix B}
	\vspace{-1mm}
	Constraints (\ref{SINR2}) and (\ref{SINR3}) is reformulated as 
	\begin{align}
		\label{SINR2-0}
		&	\begin{bmatrix}
			\eta_{c,k}-\mathcal{A}_{c,k} -\sigma_2^2 &  (\mathbf{h}_{c,k}+\mathbf{u}_{c,k}^{\rm T}\mathbf{F}_{c,k})\mathbf{W}_{c,-k} \\
			\mathbf{W}_{c,-k}^{\mathrm{H}}(\mathbf{h}_{c,k}^{\rm H}+\mathbf{F}_{c,k}^{\rm H}\mathbf{u}_{c,k})  & \mathbf{I}_{K-1}
		\end{bmatrix} \succeq \mathbf{0}, \nonumber \\
		&\Lambda_{h,c,k},~\Lambda_{F,c,k},~\forall c,~k,
	\end{align}
	\begin{align}
		\label{A_k} 
		&	\begin{bmatrix}
			\mathcal{A}_{c,k} &  \sigma_1 \mathbf{f}_{c,k}\boldsymbol{\Theta}_c \\
			\sigma_1 \boldsymbol{\Theta}_c^{\mathrm{H}}\mathbf{f}_{c,k}^{\mathrm{H}}  & \mathbf{I}_M
		\end{bmatrix} \succeq \mathbf{0}, 
		~\Lambda_{f,c,k},~\forall c,~k.
	\end{align}
	By introducing $\mathbf{h}_{c,k}=\mathbf{\widehat{h}}_{c,k}+\Delta\mathbf{h}_{c,k}$, $\mathbf{F}_{c,k}=\mathbf{\widehat{F}}_{c,k}+\Delta\mathbf{F}_{c,k}$, and $\mathbf{f}_{c,k}=\mathbf{\widehat{f}}_{c,k}+\Delta\mathbf{f}_{c,k}$ in constraint (\ref{SINR2-0}), we have
	\begin{align}
		\label{SINR2-2}
		\!\!\!\!\!\!\!\!  \mathbf{0} \preceq  & \begin{bmatrix}
			\eta_{c,k}-\mathcal{A}_{c,k} - \sigma_2^2 & \mathbf{t}_{c,k} \\
			\mathbf{t}_{c,k}^{\rm H}  & \mathbf{I}_{K-1}
		\end{bmatrix} +
		\begin{bmatrix}
			\mathbf{0} & \widetilde{\mathbf{t}}_{c,k} \\ 
			\widetilde{\mathbf{t}}_{c,k}^{\rm H}  & \mathbf{0}
		\end{bmatrix}  \nonumber \\
		\!\!\!\!\!\!\!\! \preceq &
		\begin{bmatrix}
			\mathbf{0} \\ \mathbf{W}_{c,-k}^{\rm H}
		\end{bmatrix} 
		\begin{bmatrix}
			\Delta\mathbf{h}_{c,k}^{\rm H} & \mathbf{0}
		\end{bmatrix} 
		+
		\begin{bmatrix}
			\Delta\mathbf{h}_{c,k} \\ \mathbf{0}
		\end{bmatrix}
		\begin{bmatrix}
			\mathbf{0} & \mathbf{W}_{c,-k} \\
		\end{bmatrix} \nonumber \\  
		\!\!\!\!\!\!\!\! & +
		\begin{bmatrix}
			\mathbf{u}_{c}^{\rm H} \\ \mathbf{0}
		\end{bmatrix} \Delta\mathbf{F}_{c,k}
		\begin{bmatrix}
			\mathbf{0} & \mathbf{W}_{c,-k}
		\end{bmatrix}  
		+	\begin{bmatrix}
			\mathbf{0} \\ \mathbf{W}_{c,-k}^{\rm H}
		\end{bmatrix} \Delta\mathbf{F}_{c,k}^{\rm H}
		\begin{bmatrix}
			\mathbf{u}_{c} & \mathbf{0}
		\end{bmatrix}  \nonumber \\
		\!\!\!\!\!\!\!\! &+
		\begin{bmatrix}
			\eta_{c,k}-\mathcal{A}_{c,k}-\sigma_2^2 & \mathbf{t}_{c,k} \\
			\mathbf{t}_{c,k}^{\rm H}  & \mathbf{I}_{K-1}
		\end{bmatrix},
	\end{align}
	where $\mathbf{t}_{c,k}=(\widehat{\mathbf{h}}_{c,k}+\mathbf{u}_c^{\rm H}\widehat{\mathbf{F}}_{c,k})\mathbf{W}_{c,-k}$ and $\widetilde{\mathbf{t}}_{c,k}=(\Delta\mathbf{h}_{c,k}+\mathbf{u}_c^{\rm H}\Delta\mathbf{F}_{c,k})\mathbf{W}_{c,-k}$.
	
	For constraint (\ref{A_k}), we have
	\begin{align}
		\label{A_k1}
		&  \begin{bmatrix}
			\mathcal{A}_{c,k} & \sigma_1 \mathbf{f}_{c,k}\boldsymbol{\Theta}_c \\
			\sigma_1 \boldsymbol{\Theta}_c^{\mathrm{H}}\mathbf{f}_{c,k}^{\mathrm{H}}  & \mathbf{I}_{M}
		\end{bmatrix} +
		\begin{bmatrix}
			\mathbf{0} \\ \sigma_1 \boldsymbol{\Theta}_c^{\mathrm{H}}
		\end{bmatrix} 
		\begin{bmatrix}
			\Delta\mathbf{f}_{c,k}^{\mathrm{H}}  & \mathbf{0}
		\end{bmatrix}  \nonumber \\
		&+
		\begin{bmatrix}
			\Delta\mathbf{f}_{c,k} \\ 
			\mathbf{0}
		\end{bmatrix} 
		\begin{bmatrix}
			\mathbf{0} & \sigma_1 \boldsymbol{\Theta}_c
		\end{bmatrix} 
		\succeq \mathbf{0}, ~\Lambda_{f,c,k}.
	\end{align}
	
	\vspace{-2mm}
	\section{The Derivation of (\ref{Rate})} \label{Appendix C}
	\vspace{-1mm}
	To tackle constraint (\ref{Rate}), we transform it as 
	\begin{align}
		\label{Rate1}
		&\mathrm{Pr}\Bigl\{(\widehat{\mathbf{h}}_{c,k}+\mathbf{u}_c^{\mathrm{H}}\widehat{\mathbf{F}}_{c,k})\mathbf{C}_{c,k}(\widehat{\mathbf{h}}_{c,k}^{\mathrm{H}}+\widehat{\mathbf{F}}_{c,k}^{\mathrm{H}}\mathbf{u}_c)  \nonumber \\
		&+2\Re\left\{(\widehat{\mathbf{h}}_{c,k}+\mathbf{u}_c^{\mathrm{H}}\widehat{\mathbf{F}}_{k})\mathbf{C}_{c,k} (\Delta\mathbf{h}_{c,k}^{\mathrm{H}}+\Delta\mathbf{F}_{c,k}^{\mathrm{H}}\mathbf{u}_c)\right\} \nonumber \\
		&+(\Delta\mathbf{h}_{c,k}+\mathbf{u}_c^{\mathrm{H}}\triangle\mathbf{F}_{c,k})\mathbf{C}_{c,k}(\Delta\mathbf{h}_{c,k}^{\mathrm{H}}+\Delta\mathbf{F}_{c,k}^{\mathrm{H}}\mathbf{u}_c) \nonumber \\ 
		& - \sigma_1^2\widehat{\mathbf{f}}_{c,k}\boldsymbol{\Phi}_c\widehat{\mathbf{f}}_{c,k}^{\mathrm{H}} - 2\Re\{\sigma_1^2\widehat{\mathbf{f}}_{c,k}\boldsymbol{\Phi}_c\Delta\mathbf{f}_{c,k}^{\mathrm{H}}\}\nonumber \\
		& 
		-\sigma_1^2\Delta{\mathbf{f}}_{c,k}\boldsymbol{\Phi}_c\Delta{\mathbf{f}}_{c,k}^{\mathrm{H}} -\sigma_{2}^{2} \geq 0 \Bigr\}.
	\end{align}
	Then, we recast constraint (\ref{Rate1}) as follows:
	\begin{align}
		\label{Rate1-1}
		2\Re& \left\{(\widehat{\mathbf{h}}_{c,k} \! +\! \mathbf{u}_c^{\mathrm{H}}\widehat{\mathbf{F}}_{k})\mathbf{C}_{c,k}\left(\Delta\mathbf{h}_{c,k}\! +\! \Delta\mathbf{F}_{c,k}^{\mathrm{H}}\mathbf{u}_c\right) \!-\! \sigma_1^2\widehat{\mathbf{f}}_{c,k}\boldsymbol{\Phi}_c\Delta\mathbf{f}_{c,k}^{\mathrm{H}}\right\} \nonumber \\
		= &2\Re  \{ (\widehat{\mathbf{h}}_{c,k}+\mathbf{u}_c^{\mathrm{H}}\widehat{\mathbf{F}}_{c,k}) \mathbf{C}_{c,k} \Delta\mathbf{h}_{c,k} - \sigma_1^2\widehat{\mathbf{f}}_{c,k}\boldsymbol{\Phi}_c\Delta\mathbf{f}_{c,k}^{\mathrm{H}}\nonumber \\
		&+ \mathrm{vec}^{\mathrm{T}} (\mathbf{u}_c(\widehat{\mathbf{h}}_{c,k}+\mathbf{u}_c^{\mathrm{H}}\widehat{\mathbf{F}}_{c,k})\mathbf{C}_{c,k})\mathrm{vec}(\Delta\mathbf{F}_{c,k}^{\ast}) \} \nonumber \\
		= &2\Re\{\varpi_{h,c,k}((\widehat{\mathbf{h}}_{c,k}\! +\! \mathbf{u}_c^{\mathrm{H}}\widehat{\mathbf{F}}_{c,k}) \mathbf{C}_{c,k})\mathbf{i}_{h,c,k} \!-\! \sigma_1^2\varpi_{f,c,k}\widehat{\mathbf{f}}_{c,k}\boldsymbol{\Phi}_c\mathbf{i}_{f,c,k}^{\mathrm{H}}  \nonumber \\
		& + \varpi_{F,c,k}\mathrm{vec}^{\mathrm{T}} (\mathbf{u}_c(\widehat{\mathbf{h}}_{c,k}+\mathbf{u}_c^{\mathrm{H}}\widehat{\mathbf{F}}_{c,k})\mathbf{C}_{c,k})\mathbf{i}_{F,c,k}^{\ast}  \} \nonumber \\
		=& 2\Re \{ \mathbf{e}_{c,k}^{\mathrm{H}} \widetilde{\mathbf{i}}_{c,k}\},
	\end{align}
	where $\widetilde{\mathbf{i}}_{c,k} = [\mathbf{i}_{h,c,k}^{\mathrm{H}} \quad \mathbf{i}_{F,c,k}^{\mathrm{T}} \quad \mathbf{i}_{f,c,k}]^{\mathrm{H}}$ and 
	$\mathbf{e}_{c,k} = 
	\begin{bmatrix}
		\varpi_{h,c,k}\mathbf{C}_{c,k}(\widehat{\mathbf{h}}_{c,k}^{\mathrm{H}}+\widehat{\mathbf{F}}_{c,k}^{\mathrm{H}}\mathbf{u}_c)  \\
		\varpi_{F,c,k}\mathrm{vec}^{\ast} (\mathbf{u}_c(\widehat{\mathbf{h}}_{c,k}+\mathbf{u}_c^{\mathrm{H}}\widehat{\mathbf{F}}_{c,k})\mathbf{C}_{c,k}) \\
		-\sigma_1^2\varpi_{f,c,k}\boldsymbol{\Phi}_c^{\mathrm{H}}\widehat{\mathbf{f}}_{c,k}^{\mathrm{H}}
	\end{bmatrix}$.
	
	Then we have 
	\begin{align}
		\label{Rate1-3}
		\!\! (\Delta&\mathbf{h}_{c,k}\!+\!\mathbf{u}_c^{\mathrm{H}}\triangle\mathbf{F}_{c,k})\mathbf{C}_{c,k}\left(\Delta\mathbf{h}_{c,k}^{\mathrm{H}}\!+\!\Delta\mathbf{F}_{c,k}^{\mathrm{H}}\mathbf{u}_c\right) \!-\! \sigma_1^2\Delta{\mathbf{f}}_{c,k}\boldsymbol{\Phi}_c\Delta{\mathbf{f}}_{c,k}^{\mathrm{H}} \nonumber \\
		\!\! = &\Delta\mathbf{h}_{c,k}\mathbf{C}_{c,k}\Delta\mathbf{h}_{c,k}^{\mathrm{H}} + 2\Re\{\mathbf{u}_c^{\mathrm{H}}\triangle\mathbf{F}_{c,k}\mathbf{C}_{c,k} \Delta\mathbf{h}_{c,k}^{\mathrm{H}}\}  \nonumber \\
		\!\!& +\mathbf{u}_c^{\mathrm{H}}\Delta\mathbf{F}_{c,k}\mathbf{C}_{c,k}\Delta\mathbf{F}_{c,k}^{\mathrm{H}}\mathbf{u}_c - \sigma_1^2\Delta{\mathbf{f}}_{c,k}\boldsymbol{\Phi}_c\Delta{\mathbf{f}}_{c,k}^{\mathrm{H}} \nonumber \\
		\!\! =&\varpi_{h,c,k}^2 \mathbf{i}_{h,c,k} \mathbf{C}_{c,k}\mathbf{i}_{h,c,k}^{\mathrm{H}} +2\Re\{ \Delta\mathbf{h}_{c,k} (\mathbf{C}_{c,k} \otimes \mathbf{u}_c^{\mathrm{T}}) \mathrm{vec}(\Delta\mathbf{F}_{F,c,k}^{\ast})\} \nonumber \\
		\!\! &+\mathrm{vec}^{\mathrm{T}}(\Delta\mathbf{F}_{F,c,k})(\mathbf{C}_{c,k} \otimes \boldsymbol{\Phi}_c^{\mathrm{T}}) \mathrm{vec}(\Delta\mathbf{F}_{F,c,k}^{\ast}) \nonumber \\
		\!\!&- \sigma_1^2\varpi_{f,c,k}^2 \mathbf{i}_{f,c,k}\boldsymbol{\Phi}_{c} \mathbf{i}_{f,c,k}^{\mathrm{H}} \nonumber \\
		\!\! =&\varpi_{h,c,k}^2 \mathbf{i}_{h,c,k} \mathbf{C}_{c,k}\mathbf{i}_{h,c,k}^{\mathrm{H}} + \varpi_{F,c,k}^2 \mathbf{i}_{F,c,k}^{\mathrm{T}}(\mathbf{C}_{c,k} \otimes \boldsymbol{\Phi}_c^{\mathrm{T}})\mathbf{i}_{F,c,k}^{\ast} \nonumber \\
		\!\! &+ 2\Re\{\varpi_{h,c,k} \varpi_{F,c,k} \mathbf{i}_{h,c,k}(\mathbf{C}_{c,k} \otimes \mathbf{u}_c^{\mathrm{T}}) \mathbf{i}_{F,c,k}^{\ast} \}   \nonumber \\
		\!\! &- \sigma_1^2\varpi_{f,c,k}^2 \mathbf{i}_{f,c,k}\boldsymbol{\Phi}_{c} \mathbf{i}_{f,c,k}^{\mathrm{H}} =\widetilde{\mathbf{i}}_{c,k}^{\mathrm{H}}{\mathbf{E}}_{c,k} \widetilde{\mathbf{i}}_{c,k},
	\end{align}
	where $\mathbf{E}_{c,k} = $
	\begin{align}
		\label{Rate1-4}
		& \!\!\!
		\begin{bmatrix}
			\varpi_{h,c,k}^2\mathbf{C}_{c,k} \!\!\!\!\!\!&\!\!\! \varpi_{h,c,k}\varpi_{F,c,k}(\mathbf{C}_{c,k} \otimes \mathbf{u}_c^{\mathrm{T}})  \!\!\!\!\!\!& \!\!\!\!\! \mathbf{0}_{N \times M} \\
			\varpi_{h,c,k}\varpi_{F,c,k}(\mathbf{C}_{c,k} \otimes \mathbf{u}_c^{\ast}) \!\!\!\!\!\! & \!\!\!\!\! \varpi_{F,c,k}^2(\mathbf{C}_{c,k} \otimes \boldsymbol{\Phi}_c^{\mathrm{T}}) \!\!\!\!\!\!\!\!&\!\!\!\!\!\!\!\! \mathbf{0}_{MN \times M} \\
			\mathbf{0}_{M \times N} \!\!\!\!\!\!\!\!\! &\!\!\!\!\!\!\! \mathbf{0}_{M \times MN} \!\!\!\!\!\!\!\!&\!\!\!\!\!\!\!\! -\sigma_1^2\varpi_{f,c,k}^2\boldsymbol{\Phi}_c \nonumber 
		\end{bmatrix}. 
	\end{align}
	
	\vspace{-2mm}
	\section{The Transformation of problem (\ref{P5-2})} \label{Appendix D}
	For the term ${e}_{c,k} = (\widehat{\mathbf{h}}_{c,k}+\mathbf{u}_c^{\mathrm{H}}\widehat{\mathbf{F}}_{c,k})\mathbf{C}_{c,k}(\widehat{\mathbf{h}}_{c,k}^{\mathrm{H}}+\widehat{\mathbf{F}}_{c,k}^{\mathrm{H}}\mathbf{u}_c) - \widehat{\mathbf{f}}_{c,k}\boldsymbol{\Phi}_c\widehat{\mathbf{f}}_{c,k}^{\mathrm{H}} - \sigma_2^2 $ in constraint (\ref{e_c}), we have $\bar{e}_{c,k} = {e}_{c,k} =\mathrm{Tr}(\widehat{\mathbf{H}}_{c,k}\mathbf{C}_{c,k}\widehat{\mathbf{H}}_{c,k}^{\mathrm{H}}\mathbf{V}_c) -\widehat{\mathbf{f}}_{c,k}\boldsymbol{\Phi}_c\widehat{\mathbf{f}}_{c,k}^{\mathrm{H}} - \sigma_2^2$.
	
	For $\Vert \boldsymbol{\Phi}_{c} \Vert_F^2$, by adopting the first Taylor expansion, we have $\Vert \boldsymbol{\Phi}_{c} \Vert _{F}^{2} = \Phi_c= \mathrm{Tr}(2\Re{(\boldsymbol{\Phi}_{c}^{(\tau),\mathrm{H}}\boldsymbol{\Phi}_{c})}- \boldsymbol{\Phi}_{c}^{(\tau), \mathrm{H}}\boldsymbol{\Phi}_{c}^{(\tau)})$. 
	Then for constraint (\ref{P5-2-1}), we have
	\begin{align}
		\widetilde{E}_{c,k}  =&(\varpi_{h,c,k}^{4} +2\varpi_{h,c,k}^{2}\varpi_{F,c,k}^{2}\mathrm{Tr}({\boldsymbol{\Phi}_c}) +\varpi_{F,c,k}^{4}\Phi_c)\Vert\mathbf{C}_{c,k}\Vert_{F}^{2} \nonumber \\
		&- \sigma_1^4\varpi_{f,c,k}^4 \Phi_c.
	\end{align}
	For $\Vert \widehat{\mathbf{f}}_{c,k}\boldsymbol{\Phi}_c \Vert_F^2$ in constraint (\ref{P5-2-1}), we take $ \Vert \widehat{\mathbf{f}}_{c,k}\boldsymbol{\Phi}_c \Vert_{2}^{2} = f_{c,k} = 2\Re{((\widehat{\mathbf{f}}_{c,k}\boldsymbol{\Phi}_c^{(\tau)})^{\mathrm{H}}(\widehat{\mathbf{f}}_{c,k}\boldsymbol{\Phi}_c))}- (\widehat{\mathbf{f}}_{c,k}\boldsymbol{\Phi}_c^{(\tau)})^ {\mathrm{H}}(\widehat{\mathbf{f}}_{c,k}\boldsymbol{\Phi}_c^{(\tau)})$. Then we have
	\begin{align}
		\!\!\!\!\! \widetilde{e}_{c,k} = & -\sigma_1^4\varpi_{f,c,k}^2 f_{c,k}+(\varpi_{h,c,k}^{2} \nonumber \\
		\!\!\!\!\! &+\varpi_{F,c,k}^{2}\sum\nolimits_{m} \beta_m^c)\mathrm{Tr}(\widehat{\mathbf{H}}_{c,k}\mathbf{C}_{c,k}\mathbf{C}_{c,k}^{\mathrm{H}}\widehat{\mathbf{H}}_{c,k}^{\mathrm{H}}\mathbf{V}_c). 
	\end{align}   
	
	\begin{spacing}{1}
		
	\end{spacing}
	
\end{document}